\documentclass[preprintnumbers,amsmath,amssymb,floatfix,10pt,prd,onecolumn,
superscriptaddress,nofootinbib]{revtex4}
\usepackage{latexsym}
\usepackage{epsfig}
\usepackage{epstopdf}
\usepackage{graphicx}
\usepackage{amssymb}
\usepackage{amsmath}
\usepackage{dcolumn}
\usepackage{bm}
\usepackage{color}
\usepackage{comment}
\usepackage[utf8]{inputenc}
\begin{document}

\title{Energy Constraints for Evolving Spherical and Hyperbolic Wormholes in $f(R,T)$ Gravity}

\author{M. Zubair}
\email{mzubairkk@gmail.com; drmzubair@cuilahore.edu.pk}\affiliation{Department
of Mathematics, COMSATS University Islamabad, Lahore Campus,
Lahore, Pakistan}

\author{Quratulien Muneer}
\email{anie.muneer@gmail.com}\affiliation{Department of Mathematics, COMSATS University Islamabad, Lahore Campus, Lahore, Pakistan}

\author{Saira Waheed}
\email{swaheed@pmu.edu.sa}\affiliation{Prince Mohammad Bin Fahd University, Khobar, Kingdom of Saudi Arabia}

\begin{abstract}

The primary objective of this article is to study the energy condition bounds for spherical and hyperbolic
wormholes in well-known $f(R,T)$ theory of gravity. For this purpose, we formulate the field equations for spherically and
pseudospherically geometries using anisotropic matter and linear form of generic function $f(R,T)$. By imposing different
conditions on radial and tangential pressures or by adopting some known choices for red shift and shape functions, we present the
graphical analysis of energy conditions for both spherically and pseudospherically symmetric wormholes. It is seen that energy
density for spherically symmetric wormhole is always positive for $\lambda>-4\pi$ and $\lambda<-8\pi$, while the energy conditions
for radial pressure are negative at throat. Likewise, in case of pseudospherically symmetric wormhole, it is observed that energy
density is always positive for negative $\lambda$, however conditions based on radial pressure may be positive or negative for
the considered different cases.\\

\textbf{Keywords}: Modified gravity, Wormholes, Spherically symmetric spacetime, Pseudospherically symmetric spacetime, Energy conditions.
\end{abstract}

\maketitle

\date{\today}

\section{Introduction}

General relativity (GR) embodies the conceptual ingenuity and the very essence of scientific style in its each advancement.
It provides considerable contributions at observational as well as experimental levels which range from millimeter to cosmological
scales \cite{a}. However, despite its excellence, the geometrical singularities challenge its potential and make it impotent in
certain situations. It is argued that simple GR action cannot provide complete description of accelerated expansion eras of the
universe and it is only possible by summoning new types of matter and energy components of the universe in its action \cite{S}.
To incorporate these adjustments, modified matter models or theories have been proposed in literature. In this respect,
modified theories have gained considerable attention of researchers after the revelation of accelerated expanding nature of
universe in its current state. It is worth mentioning here that although several theories like torsion based framework and its modified
versions, scalar field theories, Gauss-Bonnet gravity have been presented \cite{ND} but the natural extensions of GR namely
$f(R)$ theory \cite{D} and its extension $f(R,T)$ gravity \cite{T} (with $R$ and $T$ as Ricci scalar and trace of stress energy tensor,
respectively) have gained more attention due to their success and viability. Several works are available in literature \cite{b}-\cite{23}
which present discussions on various cosmic issues and proved as best candidates for the explanation of cosmic acceleration.

Geometrically, wormholes are tunnels type structures which join two different asymptotically flat distant universes or two
different regions of the same universe. Such tube-like structures were firstly introduced by Einstein and Rosen \cite{1*}, who
investigated exact solutions of GR, and their proposed structures were named as Einstein-Rosen bridges. Later, Morris and Thorne (M-T)
\cite{1} explored the static spherically symmetric wormhole by adopting GR gravitational framework and proposed the traversable wormhole by
describing its fundamental aspects, according to which human body can travel via wormhole tunnels. It is worthy to mention here
that wormhole geometry is based on two functions namely red shift and shape functions. It has been argued \cite{1} that for the
existence of a viable wormhole structure, these functions must satisfy some axioms, i.e., for $r_{0}\leq r < \infty$, where $r_{0}$
is the throat radius, the shape function $b(r)$ must satisfy the condition $b(r_{0})=r_{0}$ at $r = r_{0}$ and for $r >r_{0}$,
the conditions $1-\frac{b(r)}{r}> 0$ and $b^{'}(r_{0})< 1$ must hold. The wormhole spacetime geometry is asymptotically flat
if it satisfies the condition, i.e., limit $\frac{b(r)}{r} \rightarrow 0$ as $|r|$ $\rightarrow \infty$. Also, the red shift
$\phi(r)$ must be finite throughout the spacetime. In literature, much work have already been done on this topic in both GR as well
as modified gravity theories. In \cite{2}, authors have studied spherically symmetric traversable wormhole spacetime filled with
single fluid involving anisotropic pressure. Cataldo and his collaborators \cite{3} proposed the wormhole shape function defined
by $b(r)=\mu+\nu(r)$ while Samanta et al. \cite{3*} discussed the exponential form of shape function, i.e., $b(r)=\frac{r}{e^{(r-r_0 )}}$
in GR framework. Further, in another study \cite{3**}, authors have used the power law type function given by $b(r)=(r_0)^{\epsilon+1}(r)^{-\epsilon}$
by varying $\epsilon$. In \cite{3,4}, the authors have studied the existence of spherically symmetric wormholes as well as
pseudospherically symmetric spacetimes satisfying all necessary conditions for shape and red shift functions. They concluded that
the energy density becomes negative and radial pressure remains positive for hyperbolic wormhole in GR. In a recent paper \cite{4*},
Sharma and Ghosh proposed a generalized version of Ellis-Bronnikov wormholes which is embedded in a 4-dimensional wrapped
background and discussed the validity of energy constraints in GR.

Many authors have explored the wormhole solutions in modified theory including Azizi \cite{4**} who studied wormhole geometries in
the framework of $f(R,T)$ theory by considering a particular EoS for ordinary matter and concluded that modified stress-energy
is responsible for the violation of the NEC. Sahoo with his collaborators \cite{5} proposed the modeling of traversable wormholes
(a wormhole in which any body can be passed safely) in the frame work of traceless $f(R,T)$ gravity theory. They discussed
different features of energy condition bounds by considering hybrid form of shape function and a particular form of
EoS parameter. Similarly, Dixit and his collaborators \cite{5*} explored the traversable wormhole by applying logarithmic
shape function in $f(R,T)$ gravity and discussed the behavior of radial and tangential components of state along with
anisotropy parameter in describing the universe geometry. Mishra et al. \cite{7} presented the traversable wormhole models using different ansatz
of shape functions and EoS parameter in the $f(R,T)$ gravitational framework. They also examined the suitability of these models
by exploring all energy constraints. Further, Sahoo and his collaborators \cite{8} studied the wormhole solution for phantom case, i.e.,
$\omega<-1$ in $f(R,T)$ modified gravity. They adopted the EoS given by $p_r=\omega\rho$ to obtain the shape function which obeys all
its axioms and also examined corresponding energy condition bounds. Similarly, Moraes and Sahoo \cite{9} proposed the wormhole solution
by applying power law form of shape function and analyzed all energy constraints in exponential $f(R,T)$ gravity. In literature \cite{10}-\cite{20*},
wormholes have been discussed without non-exotic matter in modified theories. In this respect, Zubair et al. \cite{21} proposed wormhole
geometries by adopting simple linear as well as cubic forms of $f(R,T)$ function. They considered the non-commutative geometrical aspects
of string theory in the context of $f(R,T)$ gravity. In another paper \cite{z32}, the researchers have explored the idea of viable charged
wormhole solutions in $f(R,T)$ gravity. They have assumed simple linear generic model given by $f(R,T)=R+2T$ along with the
ordinary matter as the total pressure of anisotropic fluid. Further, the existence of static wormhole model have been explored by
utilizing different kind of shape functions in a study \cite{z33}. In another paper \cite{z34}, authors have investigated
the wormhole modeling by considering a specific general shape function in the quadratic $f(R,T)$ gravity. In a recent paper \cite{z35},
Banerjee and his collaborator adopted different strategies to construct wormhole geometry using isotropic pressure in $f(R,T)$ theory and
concluded that all energy constraints are valid for the proposed models.

Above literature motivated us to explore the spherically symmetric and pseudospherically symmetric wormhole solutions for M-T
spacetime in $f(R,T)$ theory. Here, in order to calculate the $\phi(r)$ and $b(r)$ functions, we consider the matter with EoS
parameter for $p_r$ or $p_t$ to solve $f(R,T)$ field equations. Also, we shall present the validity of energy conditions for
static spherically and psuedospherically symmetric wormhole spacetimes defined by the single perfect fluid, i.e., a source of matter
with isotropic pressure in $f(R,T)$ gravity with wide range of $\lambda$. Up to our knowledge, till now, the only wormhole solution
discussed in literature are non-asymptotically flat wormhole with isotropic pressure as pointed out in Ref. \cite{28}-\cite{30}.
In the present work, we shall discuss the asymptotically flat wormholes which may or may not satisfy all energy constraints
with a wide range of $\lambda$ for both types of wormhole spacetimes.

We have organized this paper in the following pattern. Section $\textrm{II}$ provides the basic formulation of $f(R,T)$ field
equations for static spherically symmetric spacetime and anisotropic matter. In its subsections, we shall discuss
the energy conditions graphically for different cases: firstly, wormholes with zero tidal force and filled with perfect fluid,
secondly, by taking the linear equation of state, and lastly, the analytic solution used by Tolman and static spheres filled
with isotropic perfect fluid. Similarly, in sec $\textrm{IV}$, we express the $f(R,T)$ field equations for hyperbolic symmetric
spacetime. In its subsection \textbf{A}, we discuss the pseudospherical symmetric wormhole and corresponding energy conditions
using linear equation of state. The zero tidal hyperbolic wormholes with the corresponding energy conditions will be explained
in subsection \textbf{B}. We choose non-vanishing red shift function and evaluate the more general form of shape function with
isotropic pressure and explore the energy conditions for the obtained hyperbolic wormhole in subsection \textbf{C}. Finally, we
summarize our discussion.

\section{Field Equations for Spherically Symmetric Spacetime in $f(R,T)$ and Wormholes Construction}

In this section, we shall present a brief review of $f(R,T)$ gravity and some necessary assumptions used
for this work. The $f(R,T)$ theory of gravity is defined by the total action \cite{T} given by
\begin{equation}\label{1}
S=\frac{1}{16\pi}\int f(R,T)\sqrt{-g}d^{4}x+\int L_{m}\sqrt{-g}d^{4}x,
\end{equation}
where $f(R,T)$ is an arbitrary function of Ricci scalar $R$ and the trace of energy-momentum tensor $T=g^{\mu\nu}T_{\mu\nu}$ and
$L_{m}$ represents the Lagrangian density of ordinary matter. By varying the above action with respect to metric tensor $g_{ij}$,
we have the following set of equations
\begin{eqnarray}\label{2}
8\pi T_{ij}-f_{T}(R,T)T_{ij}-f_{T}(R,T)\Theta_{ij}&=&f_{R}(R,T)R_{ij}-\frac{1}{2}f(R,T)g_{ij}+(g_{ij}\Box-\nabla_{i}\nabla_{j})f_{R}(R,T).
\end{eqnarray}
The above equation involves covariant derivative and d'Alembert operator denoted by $\nabla$ and $\Box$, respectively, where
$f_{R}(R,T)$ and $f_{T}(R,T)$ represent the function derivatives with respect to $R$ and $T$, respectively. Also, the term
$\Theta_{ij}$ is defined by
\begin{equation}\label{3}
\Theta_{ij}=\frac{g^{\alpha\beta}\delta T_{ij}}{\delta
g^{ij}}=-2T_{ij}+g_{ij}L_{m}-2g^{\alpha\beta}\frac{\partial^{2}L_{m}}{\partial g^{ij}\partial g^{\alpha\beta}},
\end{equation}
where $T_{ij}=diag(\rho,~-p_r,~-p_t,~-p_t)$, with $\rho$ as the matter energy density while $p_r$ and $p_t$ denote the radial
and transverse components of pressure. Also, $L_{m}=-\emph{P}$, where $\emph{P}=\frac{(p_r+2p_t)}{3}$ is the total pressure.
As a result, Eq.(\ref{3}) takes the form given by $\Theta_{ij}=-2T_{ij}-\emph{P} g_{ij}$. For the sake of simplicity in
calculations, let us consider the linear form of generic function $f(R,T)$ given by $f(R,T)=R+\lambda T$ and
consequently, the field equations (\ref{2}) can be re-written as
\begin{equation}\label{4}
R_{ij}-\frac{1}{2}Rg_{ij} = (8\pi+\lambda)T_{ij} + \frac{\lambda}{2}(\rho-\emph{P})g_{ij},
\end{equation}
where $\lambda$ is an arbitrary constant.

In \cite{1}, Morris and Thorne introduced the wormhole geometry given by the following metric:
\begin{equation}\label{5}
ds^2=e^{2\Phi(r)}dt^2-\frac{dr^2}{1-b(r)/r}-r^{2}(d\theta^{2}+sin^{2}\theta d\Phi^{2}),
\end{equation}
where the functions $\Phi(r)$ and $b(r)$ refer to red shift and shape functions, respectively and it is
worthy to mention here that the essential characteristics of a viable wormhole geometry are encoded through
these functions. Thus in order to construct the wormhole geometry, these functions must satisfy the constraints
which are presented by Morris and Thorne in Ref. \cite{1,30}. In the present work, we assume $\phi(r)$ without
horizons and hence it must be finite everywhere. Assuming single imperfect fluid and using Eqs.(\ref{4}) and (\ref{5}), we have
\begin{eqnarray}\label{6}
&&\frac{b'}{r^2}=(8\pi+\lambda)\rho + \frac{\lambda}{2}(\rho-\frac{p_{r}+2p_{t}}{3}),\\\label{6*}
&&\frac{1}{r}\bigg[\frac{b}{r^2}+2 \phi'(\frac{b}{r}-1)\bigg]=(8\pi+\lambda)(-p_{r}) + \frac{\lambda}{2}(\rho-\frac{p_{r}+2p_{t}}{3}),\\\label{6**}
&&\frac{1}{2r}\bigg[\frac{1}{r}(\phi' b + b'-\frac{b}{r})+2(\phi''+(\phi')^{2})b- \phi'(2-b')\bigg]=(8\pi+\lambda)(-p_{t}).
+\frac{\lambda}{2}(\rho-\frac{p_{r}+2p_{t}}{3}).
\end{eqnarray}
The above equations can be re-arranged to find the explicit expressions of $\rho,~ p_{r}$ and $p_{t}$ which can be written as
\begin{eqnarray}\label{7}
\rho&=&\frac{1}{12 r^{2}(4\pi+\lambda)(8\pi+ \lambda)}\bigg[b'(48\pi+\lambda(8-r\phi')+\lambda(2r(2\phi'+r(\phi')^{2}
+r\phi''))-b(3\phi'+2r(\phi')^{2}+2r\phi'')\bigg],\\\nonumber
p_{r}&=&\frac{1}{12 r^{3}(4\pi+\lambda)(8\pi+ \lambda)}\bigg[r(b'\lambda(4+r\phi^{'})-2r(-48\pi\phi^{'}-10\lambda \phi^{'}
+r\lambda(\phi^{'})^2+r\lambda \phi^{''}))-b(48\pi(1+2r \phi^{'})\\\label{7*}
&+&\lambda(12+21r\phi^{'}-2r^2((\phi')^2+ \phi^{''})))\bigg],\\\nonumber
p_{t}&=&\frac{1}{12 r^{3}(4\pi+\lambda)(8\pi+ \lambda)}\bigg[-b^{'}\bigg[24\pi(-1 + r \phi^{'}+2r^{2}(\phi^{'})^{2}+\phi^{''}))+\lambda(-6+3r\phi^{'}+10r^{2}(\phi^{'})^{2}+\phi^{''}))\bigg]\\\label{7**} &+&r\bigg[-b^{'}(24\pi(1+r\phi^{'})+\lambda(2+5r\phi^{'}))+2r(24\pi(\phi^{'}+r(\phi^{'})^{2}+r\phi^{''})+
\lambda(4\phi^{'}+5r((\phi^{'})^{2}+\phi^{''})))\bigg]\bigg],
\end{eqnarray}
where prime denotes the derivative with respect to radial coordinate. Also, from the conservation equation $T^{ij}_{;j}=0$,
the hydrostatic equilibrium equation for the ordinary matter in the interior of wormhole can be defined as
\begin{eqnarray}\label{8}
p_{r}^{'}+\phi^{'}(\rho+p_{r})=\frac{\lambda}{(16\pi+2\lambda)}(\rho^{'}-\frac{p_{r}^{'}+2p_{t}^{'}}{3}).
\end{eqnarray}
It is obvious that for isotropic fluid, the condition $p_{r}=p_{t}$ must be imposed. Using this condition, we can obtain a
differential equation connecting both redsihft and shape functions given by
\begin{eqnarray}\label{9}
\phi^{''}+\phi^{'^{2}}-\frac{(b^{'}r+2r-3b)\phi^{'}}{2r(r-b)}=\frac{b^{'}r-3b}{2r^{2}(r-b)}.
\end{eqnarray}
The above differential equation (DE) can be solved for one of these two unknowns, for example, one can solve it for
$b(r)$ by choosing a specific known form of function $\phi(r)$. Consequently, it can be written as
\begin{eqnarray}\label{10}
b(r)=\bigg(\int\frac{2r(-\phi^{'}+r\phi^{'2}+r\phi^{''})e^{\int\frac{2r^{2}(\phi^{'2}+\phi^{''})-3r\phi^{'}-3}{r(1+r\phi^{'})}dr}}{1+r\phi^{'}}dr+c\bigg)
\times e^{-\int\frac{2r^{2}(\phi^{'2}+\phi^{''})-3r\phi^{'}-3}{r(1+r\phi^{'})}dr},
\end{eqnarray}
where $c$ is an integration constant. It is interesting to point out that the $f(R,T)$ field equations has been reduced
to two independent differential equations (DEs) given by Eqs.(\ref{7}) and (\ref{9}) which involve four unknowns namely $\phi(r),~ \rho(r),~ b(r)$
and $p(r)$ under the assumption of isotropic fluid. Therefore, in order to obtain exact solutions, we have to consider some known
forms of two functions and evaluate the others. In the upcoming subsections, we shall present the wormholes solutions by
considering some interesting cases and also evaluate the energy constraint bounds in each case.

\subsection{Energy conditions}
Energy constraints play an important role in the framework of GR as well as alternative gravity theories for exploring
the physical viability of proposed models. It is argued that the set of four energy constraints can be formulated by using
the well-known Raychaudhuri's equations and these bounds were firstly developed in GR \cite{Ray} and then extended to modified gravity theories.
These energy constraints are known as the null energy condition (NEC), the weak energy condition (WEC), the dominant energy condition (DEC)
and the strong energy condition (SEC). The Raychaudhuri's equations for congruence of geodesic with time-like and null-like vectors are given by \cite{Ray1}
\begin{eqnarray}\nonumber
\frac{d\theta}{d\tau}&=&\frac{-\theta^2}{3}-\sigma_{ij}\sigma^{ij}+\omega_{ij}\omega^{ij}-R_{ij}u^iu^j,\\\nonumber
\frac{d\theta}{d\tau}&=&\frac{-\theta^2}{3}-\sigma_{ij}\sigma^{ij}+\omega_{ij}\omega^{ij}-R_{ij}k^ik^j,
\end{eqnarray}
where $\sigma_{ij},~\omega_{ij}$ and $R_{ij}$ are the shear tensor, rotation and Ricci tensor, respectively, while $u^i$ represents the time-like
vector and $u^j$ indicates the null-like geodesics in congruence. It will converge when $\theta<0$ which further leads to $\frac{d\theta}{d\tau}<0$.
By ignoring second order terms and integrating above equation, we obtain $\theta=-\tau R_{ij}u^iu^j$ and $\theta=-\tau R_{ij}k^ik^j$, where
$\sigma_{ij}\sigma^{ij}\geq0$ and $\omega=0$. Using these conditions, one can define the following inequalities
\begin{eqnarray}\nonumber
 R_{ij}u^iu^j\geq 0,~~~~~~~~~~~~~~~~~~~R_{ij}k^ik^j\geq 0
\end{eqnarray}
which can be re-written as the linear combination of energy momentum tensor and its trace by applying field equations as follows
\begin{eqnarray}\nonumber
(T_{ij}-\frac{Tg_{ij}}{2})u^iu^j\geq0,~~~~~~~~~~T_{ij}k^ik^j\geq0.
\end{eqnarray}
Now, by using above inequalities, we can define the energy constraints as WEC: $\rho=T_{ij}u^{i}u^j\geq0$, DEC: $T^{00}\geq|T^{ij}|$, i.e.,
in any orthonormal basis, the energy dominates as compared to any other components of $T^{ij}$, while
NEC: $T_{ij}k^ik^j\geq 0$ and SEC: $(T_{ij}-\frac{Tg_{ij}}{2})u^iu^j\geq0$. The NEC and SEC come from Raychaudhuri equation
which depends on the energy momentum tensor and also on alternative gravity theories, while the WEC and DEC only depend on the energy-momentum tensor.
In upcoming sections, we shall discuss the energy condition bounds for wormhole models by using these constraints.

\subsection{Wormhole with Isotropic Pressure using Zero-tidal Force Condition in $f(R,T)$ Gravity}

In this section, we shall discuss the existence of wormhole geometries by taking two cases into account. In the first case,
we consider the zero-tidal-force wormhole in which $\phi(r)=\phi_{0}$, where $\phi_0$ is a constant and consequently, Eq.(\ref{10})
leads to the form of shape function given by $b(r)=cr^{3}$. By applying the initial condition: $b(r_{0})=r_{0}$, one can determine
the value of constant $c$ and finally, the shape function takes the form as $b(r)=\frac{r^{3}}{r_0^2}$. %Thus, the behavior of shape function is shown in Fig. \ref{fig1}.
 %\begin{figure}
%\centering
%%\includegraphics[width=0.46\textwidth]{Fig1.eps}
%\includegraphics[width=0.46\textwidth]{Fig2.eps}
%\caption{\scriptsize{Evolution of energy constraints. In the right panel, $\rho$ (red dashes), $\rho+p$ (blue dashes), $\rho-p$ (gray dashes) and $\rho+3p$ (black) where chosen parameter is $r_{0}=1$.}}
%\label{fig1}
%\end{figure}
%of shape function i.e., $b(r)$ which satisfies all axioms and  In the left panel, shape function (blue dashes), $b'(r)$ (red dashes), $b(r)-r$ (gray dashes) and $\frac{b(r)}{r}$ (black line).
Consequently, the wormhole spacetime turns out to be
\begin{eqnarray}\label{11}
ds^{2}=dt^{2}-\frac{dr^{2}}{1-(r/r_{0})^{2}}-r^{2}(d\theta^{2}+sin^{2}\theta d\Phi^{2}).
\end{eqnarray}
It is worthy to mention here that the above spacetime represents a metric with constant curvature. It is easy to compute the
values of density and pressure and are given by $\rho=\frac{2(6\pi+\lambda)}{r^2_{0}(4\pi+\lambda)(8\pi+\lambda)}$
and $p=-\frac{4\pi}{r^2_{0}(4\pi+\lambda)(8\pi+\lambda)}$. Consequently, $\rho$ and $p$ are related by the expression
$p=-\rho\frac{ 2\pi}{(6\pi+\lambda)}$.Also, $b'(r_0)=3\nless 1$ which implies that flaring out condition is not satisfied,
therefore, no zero tidal force wormhole can be sustained in the present configuration.
%The behavior of $\rho$ and all energy conditions are shown in Fig. \ref{fig1}.
%The graphs represent that all energy conditions are valid for $\lambda > -4\pi$.
%at $r_{0} = 18$.
%since solution exists only for $r_{0}\geq r$.

\begin{figure}
\centering
\includegraphics[width=0.46\textwidth]{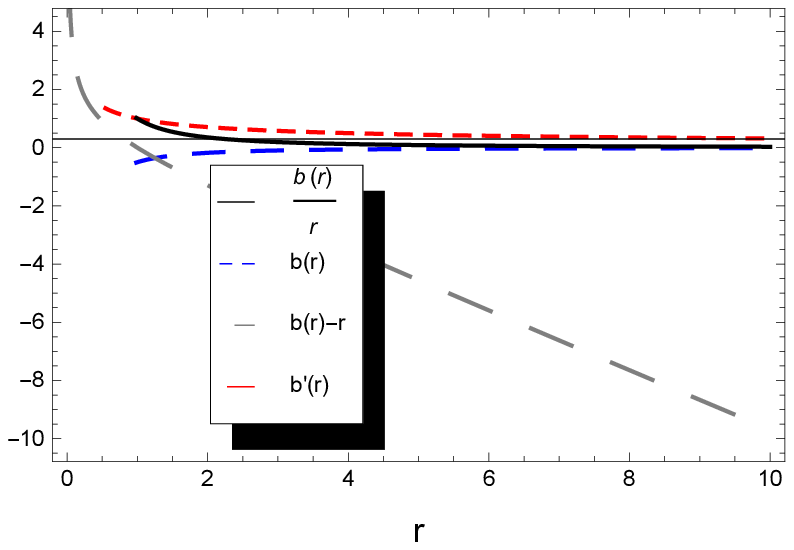}
\includegraphics[width=0.46\textwidth]{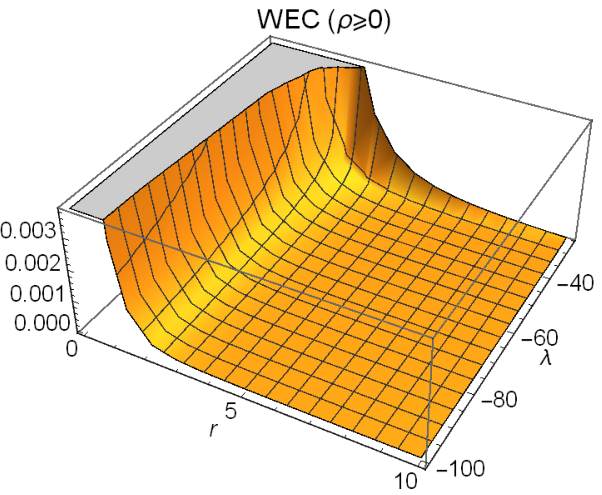}
\caption{\scriptsize{Evolution of shape function $b(r)$ satisfying the conditions and energy density versus $r$. In the left panel,
$b(r)$ (blue), $b'(r)$ (red), $b(r)-r$ (gray) and $\frac{b(r)}{r}$ (black). In the right panel, the plot indicates the behavior of
energy density ($\rho$), where the chosen parameter are $r_0=1$ and $\alpha=-.5$.}}
\label{fig2}
\end{figure}

\begin{figure}
\centering
\includegraphics[width=0.46\textwidth]{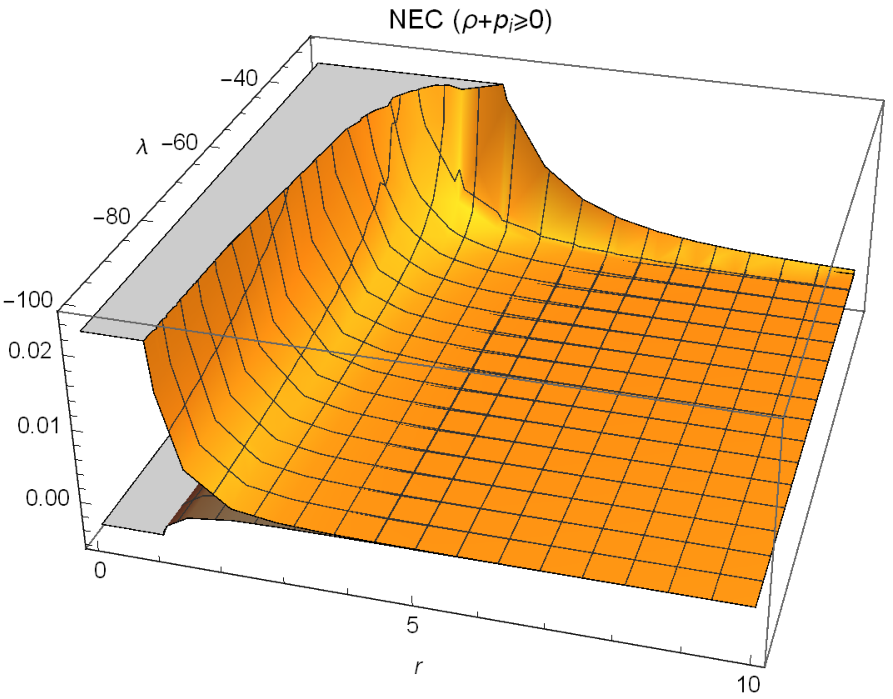}
\includegraphics[width=0.46\textwidth]{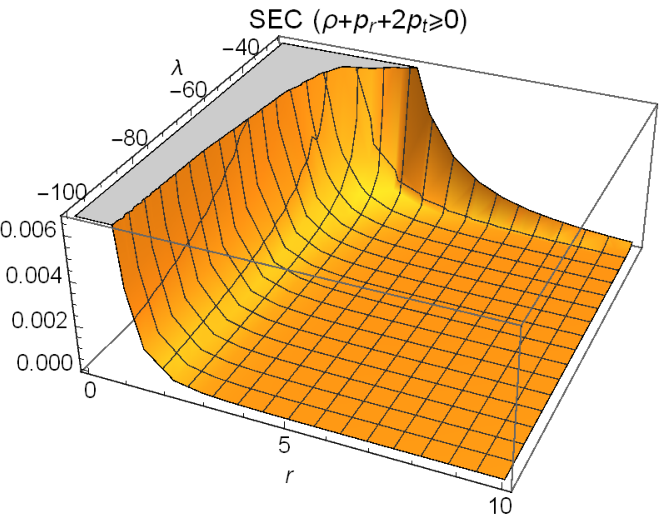}
\caption{\scriptsize{Evolution of NEC ($\rho+p_i$) and SEC ($\rho+p_r+2p_t$) versus $r$, $i=r,t$. In the left panel, the yellow plot
indicates the behavior of NEC ($\rho+p_r$) and gray plot shows behavior of NEC ($\rho+p_t$) while in the right panel, the plot
indicates the behavior of SEC where the free parameters are fixed as $r_0=1$ and $\alpha=-.5$.}}
\label{fig3}
\end{figure}

\begin{figure}
\centering
\includegraphics[width=0.46\textwidth]{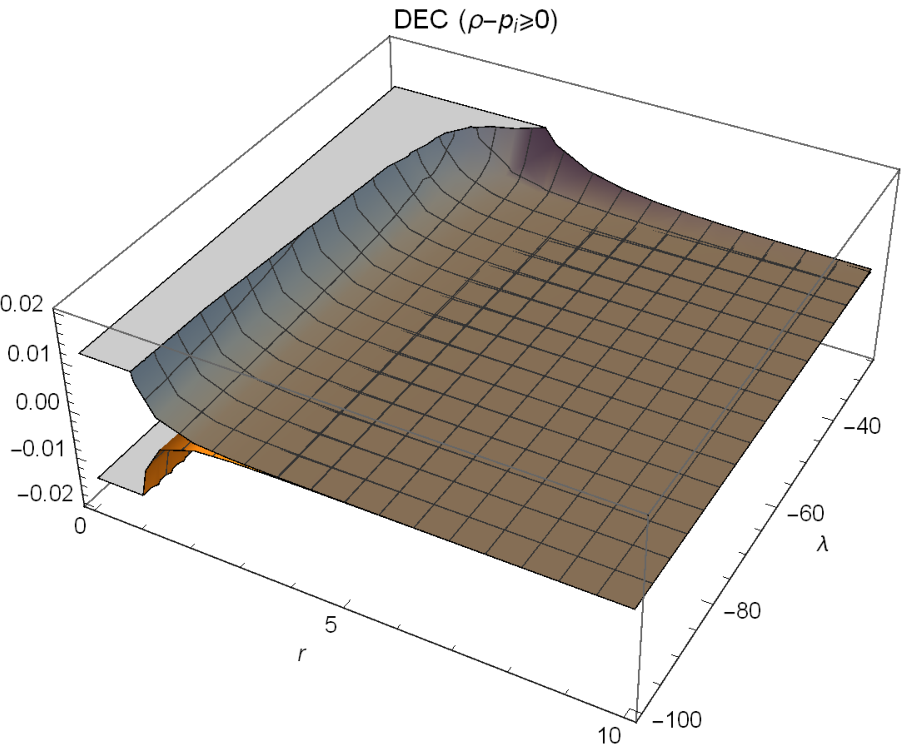}
\includegraphics[width=0.46\textwidth]{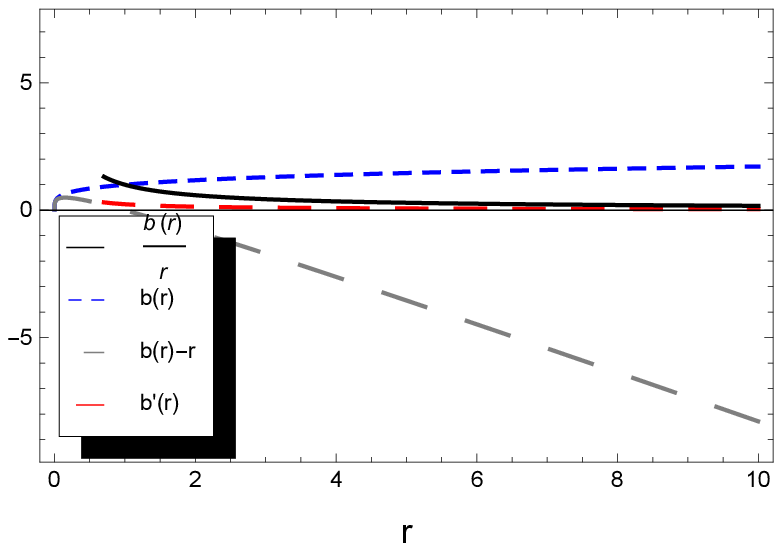}
\caption{\scriptsize{Evolution of DEC ($\rho-p_i$) and properties of shape function given by Eq.(\ref{16}) versus $r$. In the left panel,
the yellow plot indicates the behavior of DEC ($\rho-p_r$) while gray plot represents the behavior of DEC ($\rho-p_t$), where $r_0=1$ and
$\alpha=-.5$. In the right panel, shape function (blue), $b'(r)$ (red), $b(r)-r$ (gray) and $\frac{b(r)}{r}$ (black), where the chosen
parameters are $r_0=1$, $\lambda=10$ and $\omega=-5$.}}
\label{fig4}
\end{figure}

In the second case, we consider the constant red shift function and a general form of shape function given by
$b(r)=r_0(\frac{r}{r_0})^{\alpha}$, then M-T wormhole metric takes the form:
\begin{eqnarray}\label{12}
ds^{2}=dt^{2}-\frac{dr^{2}}{1-(r/r_{0})^{\alpha-1}}-r^{2}(d\theta^{2}+sin^{2}\theta d\phi^{2}),
\end{eqnarray}
where $\alpha<0$. In this case, Eqs.(\ref{7})-(\ref{7**}) lead to the following forms of energy density and pressures
\begin{eqnarray}\nonumber
\rho&=&\frac{2(1+\alpha)(\frac{r}{r_0})^\alpha(6\pi+\lambda)}{3r^2(4\pi+\lambda)(6\pi+\lambda)},\\\nonumber
p_r&=&\frac{(-12\pi+(-2+\alpha)\lambda)(\frac{r}{r_0})^\alpha}{3r^2(4\pi+\lambda)(6\pi+\lambda)}\\\nonumber,
p_t&=&-\frac{(-2\lambda+\alpha(12\pi+\lambda))(\frac{r}{r_0})^\alpha}{6r^2(4\pi+\lambda)(6\pi+\lambda)}.
\end{eqnarray}
For a viable wormhole geometry, the shape function shows increasing behavior and satisfies the conditions given by $b(r)-r<0,~ b'(r)<1$ as
well as asymptotically flat condition ($\frac{b(r)}{r}\rightarrow 0, ~r\rightarrow\infty$). It is easy to check that all of
these conditions are satisfied as shown in the left graph of Figure \ref{fig2}. Next we explore the validity of energy conditions
of the wormhole matter contents versus radial coordinate.
%These constraints are named as the weak (WEC), the null (NEC), the strong (SEC)
%and dominant energy conditions (DEC).
Mathematically, these constraints are defined as WEC ($\rho\geq 0$), NEC ($\rho+p_i\geq 0$, $i=r,t$
represents the radial and tangential pressures), SEC ($\rho+p_r+2p_t\geq0$) and DEC ($\rho-p_r, ~\rho-p_t$). The right graph of
Figure \ref{fig2} shows the graphical behavior of $\rho$ which represents that wormhole geometry satisfies the WEC for $\lambda<-8\pi$.
From Figure \ref{fig3}, it is observed that the expressions $\rho+p_i, ~i=r$ and $\rho+p_r+2p_t$ attain positive values for
$\lambda<-8\pi$, while $\rho+p_t$ exhibits negative behavior for $\lambda<-8\pi$. Thus, only NEC for radial pressure and SEC
are satisfied for $\lambda<-8\pi$. Next the graphical behavior of $\rho-p_r$ and $\rho-p_t$ is provided in Figure \ref{fig4} which indicates
that the DEC with tangential pressure is valid for $\lambda<-8\pi$, whereas DEC with radial pressure violates for $\lambda<-8\pi$.

\subsection{Spherical Wormholes with Equation of State in $f(R,T)$ Gravity}

\begin{figure}
\centering
\includegraphics[width=0.46\textwidth]{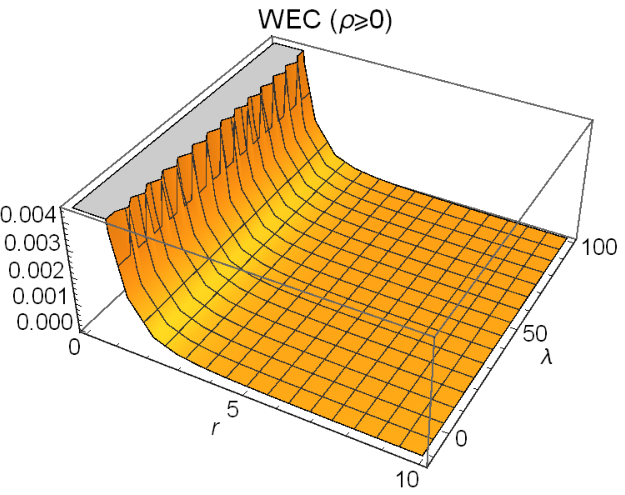}
\includegraphics[width=0.46\textwidth]{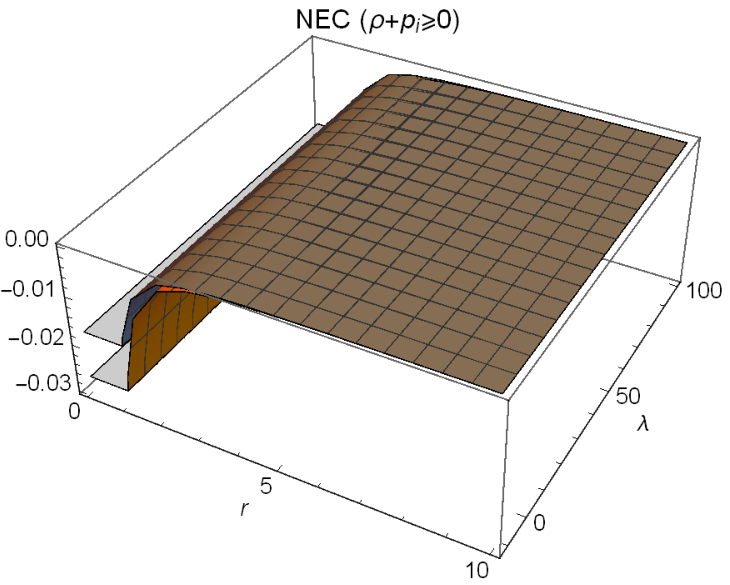}
\caption{\scriptsize{Evolution of energy density $\rho$ and NEC ($\rho+p_i,~ i=r,t$) versus $r$.
In the left panel, the plot indicates the behavior of $\rho$ and in right panel, the gray plot indicates the
behavior of $\rho+p_r$, while orange plot indicates the behavior of ($\rho+p_t$). Here we fixed $r_0=1$ and $\omega=-5$.}}
\label{fig5}
\end{figure}

\begin{figure}
\centering
\includegraphics[width=0.46\textwidth]{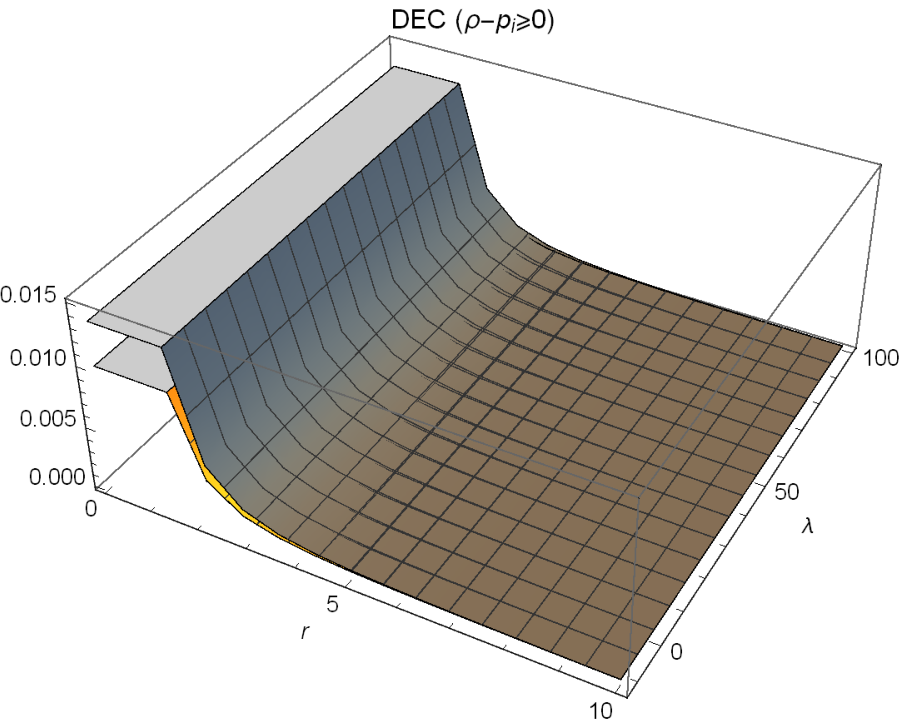}
\includegraphics[width=0.46\textwidth]{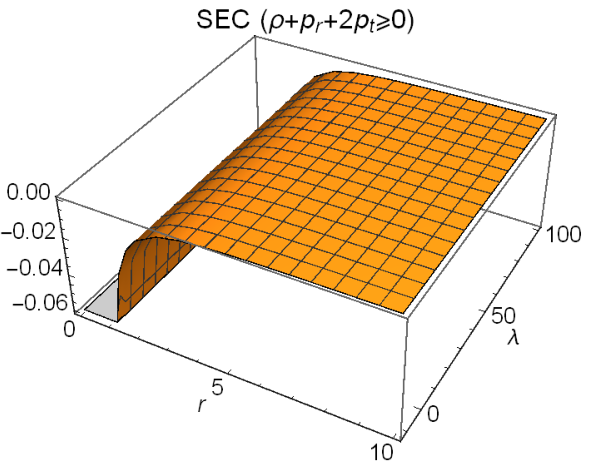}
\caption{\scriptsize{Evolution of DEC ($\rho-p_i,~ i=r,t$) and SEC ($\rho+p_r+2p_t$) versus $r$. In the left panel, the
orange plot indicates the behavior of $\rho-p_r$, while gray plot indicates the behavior of $\rho-p_t$ and in right panel,
the plot indicates the behavior of $\rho+p_r+2p_t$.}}
\label{fig6}
\end{figure}

In this subsection, we will explore the wormhole construction by considering two interesting cases of EoS parameter.
In the first place, we consider the linear equation of state along with constant red shift function and evaluate the shape
function. The linear equation of state is defined as
\begin{eqnarray}\label{14}
p_r=\omega\rho.
\end{eqnarray}
Using the above EoS along with the Eqs.(\ref{7}) and (\ref{7*}), we obtain the following differential equation:
\begin{eqnarray}\nonumber
\frac{1}{r(4\pi+\lambda)(8\pi+\lambda)}(b (48 \pi (1 + 2 r \phi^{'})-\lambda (-12 + 3 r (-7 + \omega)\phi^{'}
+ 2 r^2 (1 + \omega) (\phi^{'^2} + \phi^{''})))+\\\label{15} r (b^{'} (48 \pi \omega - \lambda (4 - 8 \omega + r \phi^{'}
+ r \omega \phi^{'})) +  2 r (-48 \pi \phi^{'} + \lambda (2 (-5 + \omega) \phi^{'} + r (1 + \omega) (\phi^{'})^2
+ r (1 + \omega) \phi^{''}))))=0.
\end{eqnarray}
For zero tidal force wormhole, i.e., $\phi(r)=0$, the above differential equation leads to the following form of shape function:
\begin{eqnarray}\label{16}
b(r)= Cr^{-\frac{3 (4 \pi + \lambda)}{12 \pi \omega + (-1 + 2 \omega) \lambda}},
\end{eqnarray}
where C is an integration constant and consequently, metric describing the wormhole geometry, energy density and tangential
pressure take the form given by
\begin{eqnarray}\nonumber
ds^{2}&=&dt^{2}-\frac{dr^{2}}{(r/r_{0})^{-\frac{(1+\omega)(12\pi+2\lambda)}{12\pi \omega-\lambda+2\omega\lambda}}-1}
-r^{2}(d\theta^{2}+sin^{2}\theta d\Phi^{2}),\\\nonumber
\rho&=&-\frac{2(6\pi+\lambda)(\frac{r}{r_0})^{-3\frac{4\pi+12\pi\omega+2\omega\lambda}{12\pi\omega-\lambda+2\omega\lambda}}}
{r_0^{2}(8\pi+\lambda)(12\pi\omega-\lambda+2\omega\lambda)},\\\nonumber
p_t&=&-\frac{(6 \pi (3 \omega + 2) - (1 - 3 \omega) \lambda)
(\frac{r}{r_0})^{-3\frac{4\pi+12\pi\omega+2\omega\lambda}{12\pi\omega-\lambda+2\omega\lambda}}}{3r_0^{2}(8\pi+\lambda)
(12\pi\omega-\lambda+2\omega\lambda)}.
\end{eqnarray}
Now we will discuss the graphical behavior of $b(r)$ defined by Eq.(\ref{16}) and check if it satisfies all axioms
for the existence of wormhole solution as discussed in the previous section. From the right plot of Figure \ref{fig4},
it is clear that the obtained shape function exhibits physically viable behavior. Further, we plot the
energy conditions of wormhole matter content versus radial coordinate. It is seen from the graphs of Figure \ref{fig5} that
the energy density remains positive while the expressions $\rho+p_r$ and $\rho+p_t$ attain negative values for $\lambda>-4\pi$
and thus, the NEC is not satisfied for this case. Also, from the graphs of Figure \ref{fig6}, it is easy to check that expressions of
DEC ($\rho-p_i$, $i=r,t$) indicate positive behavior while SEC takes negative values for $\lambda>-4\pi$. Therefore, it can be
concluded that DEC is satisfied while SEC violates in this case.

In the second case, we suppose non-vanishing red shift function, i.e., $\phi(r)\neq$ constant. For wormhole construction, we shall
construct a master differential equation for the shape function by using linear equation of state given by
\begin{eqnarray}\label{18}
p_r=p_t=\omega \rho,
\end{eqnarray}
where $\omega$ is a constant. Thus from Eq.(\ref{9}), it can be written as
\begin{eqnarray}\label{19}
\omega \rho' + \phi'(1+\omega)\rho = \frac{\lambda (1-\omega)\rho'}{16\pi+2\lambda}
\end{eqnarray}
and consequently, the energy density $\rho$ takes the form as follows
\begin{eqnarray}\label{20}
\rho(r)=c e^{-\frac{(1+\omega)(16\pi+2\lambda)\phi(r)}{16\pi\omega+3\lambda\omega-\lambda}},
\end{eqnarray}
where $c$ is an integration constant. It is interesting to mention here that if we consider vanishing red shift function, then
we shall obtain $\rho$ as a constant and this case we have already discussed in the previous subsection. From Eqs.(\ref{7}), (\ref{9})
and (\ref{18}), we obtain a differential equation for $b(r)$ as
\begin{eqnarray}\nonumber
&&(6-3r-r^2)(-16\pi+\lambda(-3+\omega))^2b^2+2r^2\bigg(-(16\pi-\lambda(-3+\omega))(16\pi(-1+r+2\omega)-\lambda(1+r(-3+\omega)\\\nonumber
&&+5\omega))b' + (-128\pi^2\omega(\omega-1)+\lambda^2(1 - 2 \omega - 3 \omega^2)+8 \pi\lambda (1 - 2 \omega + 5 \omega^2))b'^2+
r(256 \pi^2 \omega - 16 \lambda \pi (-1 + \omega^2) \\\nonumber &&+ \lambda^2 (3 - 10 \omega + 3 \omega^2)) b''\bigg)
+r(16 \pi - \lambda (-3 + \omega)) b ((16 \pi (-2 + r + 3 \omega + 2 r \omega)-\lambda (3 - 5 r + 7 \omega+ 7 r \omega))b'\\\label{20}
&&-2(-(-3 + 2 r) (16 \pi + 3 \lambda - \lambda\omega) -r (\lambda + 16 \omega\pi - 3 \omega\lambda)b'')))=0.
\end{eqnarray}
It is very difficult to find the analytic solution of above differential equation without assuming any condition. However,
one may find the solution of this equation by taking some assumptions into account. Since first term of Eq.(\ref{20}) vanishes
when $\omega=\frac{16\pi+3\lambda}{\lambda}$, $\forall$ $\lambda$, so by taking this condition, it is easy to obtain the solution:
$b(r)= A,~ e^{2\phi(r)}=1-\frac{A}{r}$ and $\rho=p=0$ which represents the Schwarzschild solution. It is worthy to mention here
that Schwarzschild solution describes the non-traversable wormhole \cite{30}. Next we consider the power law form of shape function
given by $b(r)=\frac{A}{r^n}$ which is a viable choice as it ensures the validity of M-T constraint given by $\frac{b(r)}{r}\leq 1$
for $n>-1$ and $r\rightarrow \infty$. On substitution of this choice $b(r)=\frac{A}{r^n}$ into Eq.(\ref{20}), it is easy to check that
this equation will be satisfied when the constraints $n=-3$ or -$\frac{3}{2}$ hold (for $\omega=\frac{16\pi+3\lambda}{\lambda}$ and
$\lambda=\pm 4\sqrt{2}\pi$).

\subsection{Spherical Wormhole with $e^{\phi(r)}$ = $(\frac{r}{r_{0}})^{\beta}$ in $f(R,T)$ Gravity}

In this subsection, we shall study the spherical wormhole solutions by imposing the most general form of red shift function
given by $e^{\phi(r)}=(\frac{r}{r_{0}})^{\beta}$, where $\beta$ is any arbitrary constant and try to explore the corresponding
form of shape function. By inserting $e^{\phi(r)}=(\frac{r}{r_{0}})^{\beta}$ in Eq.(\ref{9}), we obtain the shape function given
by
\begin{eqnarray}\label{21}
b(r)=\frac{\beta^{2}-2\beta}{\beta(\beta-2)-1}r - cr^{-\frac{2\beta^{2}-5\beta-3}{1+\beta}},
\end{eqnarray}
where $c$ is an integration constant.
\begin{figure}
\centering
\includegraphics[width=0.46\textwidth]{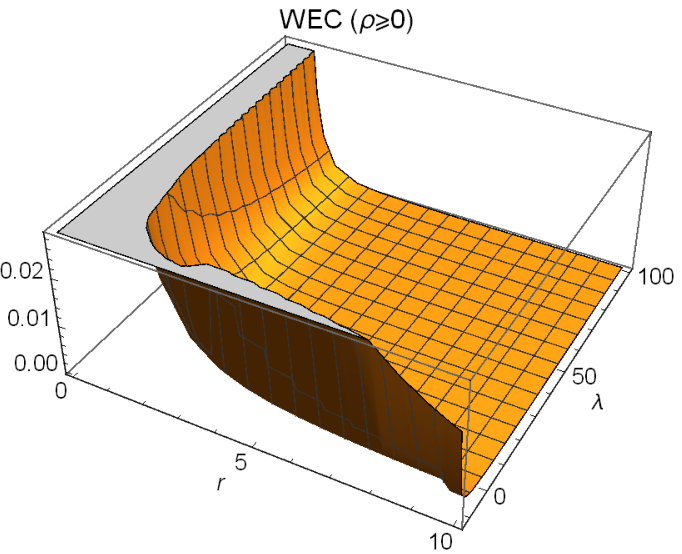}
\includegraphics[width=0.46\textwidth]{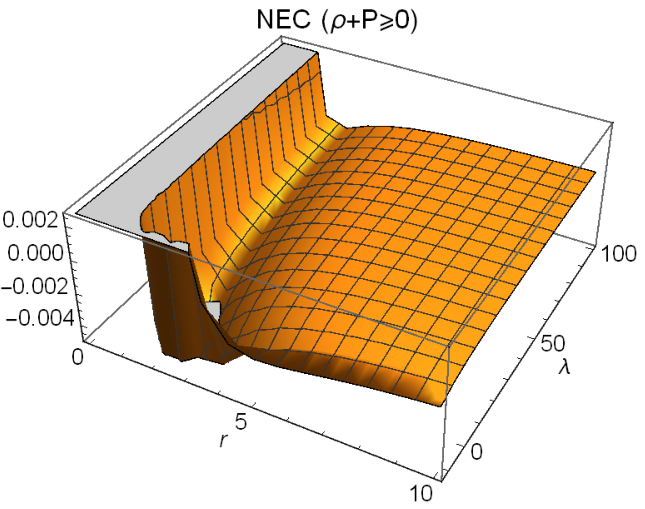}
\caption{\scriptsize{Evolution of energy density versus $r$ and NEC. The left plot indicates the behavior of energy density.
In the right panel, the gray plot indicates the behavior of $\rho+p_r$ and orange plot shows the behavior of $\rho+p_t$ for
$r_0=1$ and $\beta=.6$.}}
\label{fig7}
\end{figure}
\begin{figure}
\centering
\includegraphics[width=0.46\textwidth]{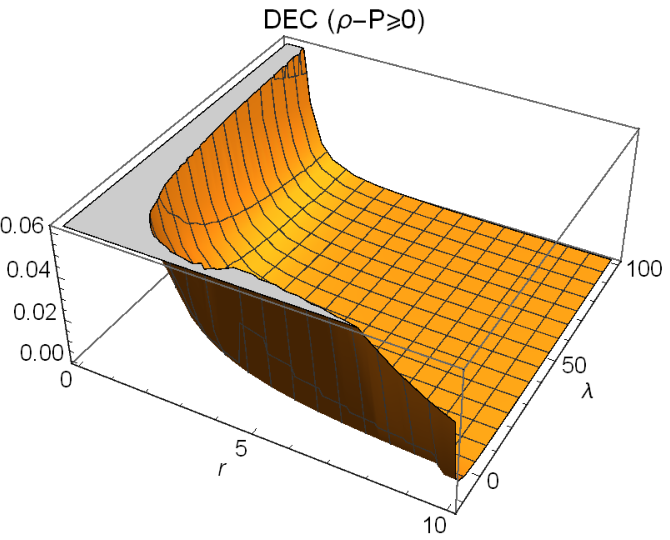}
\includegraphics[width=0.46\textwidth]{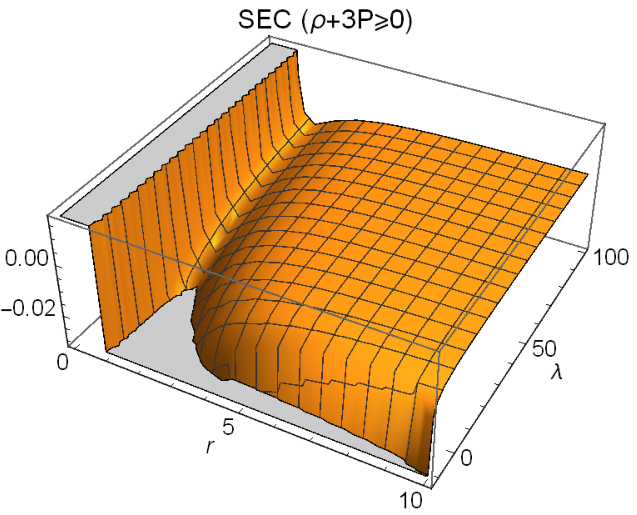}
\caption{\scriptsize{Evolution of SEC ($\rho+p_r+2p_t$) and DEC ($\rho-p_i$) versus $r$. In the left panel, the plot indicates
the behavior of SEC and in the right panel, the plot indicates the behavior of DEC, where $\rho-p_r$ (gray) and $\rho-p_t$ (orange)
for $r_0=1$ and $\beta=.6$.}}
\label{fig8}
\end{figure}
By considering the wormhole condition $b(r_{0})=r_{0}$ in Eq.(\ref{21}), we obtain the following expression for wormhole metric,
$\rho$ and $p$:
\begin{eqnarray}\label{m}
ds^2&=&(\frac{r}{r_0})^{2\beta}dt^2-\frac{dr^2}{1-\frac{\beta^2-2\beta-2}{\beta^2-2\beta-1}-c(\frac{r}{r_0})^{-\frac{2(\beta^2-2\beta-1)}{1+\beta}}}
-r^2d\Omega^2,
\end{eqnarray}
\begin{eqnarray}\nonumber
\rho=\frac{1}{12 r^{2}(4\pi+\lambda)(8\pi+ \lambda)}\bigg[\frac{6\beta(8\pi(-2+\beta)+(-3+\beta)\lambda)}{\beta(\beta-2)-1}+
\frac{12(\frac{r}{r_{0}})^{-2\frac{\beta(\beta-2)-1}{1+\beta}}(1+2\beta)(4\pi(-3+\beta)+(-2+\beta)\lambda)}{(1+\beta)(\beta(\beta-2)-1)}\bigg],
\end{eqnarray}
\begin{eqnarray}\nonumber
p=\frac{1}{12 r^{2}(4\pi+\lambda)(8\pi+ \lambda)}\bigg[-\frac{6\beta(8\pi\beta+\lambda+\lambda\beta)}{\beta(\beta-2)-1}+
\frac{12(\frac{r}{r_{0}})^{-2\frac{\beta(\beta-2)-1}{1+\beta}}(1+2\beta)(4\pi(1+\beta)+\beta\lambda)}{(1+\beta)(\beta(\beta-2)-1)}\bigg].
\end{eqnarray}
After the re-scaling $\zeta^{2}=(\beta(\beta-2)-1)r^{2}$, the metric (\ref{m}) can be written as
\begin{eqnarray}\label{25}
ds^{2} = (\frac{\zeta}{\zeta_{0}})^{2\beta}dt^{2}-\frac{d\zeta^2}{(\frac{\zeta}{\zeta_{0}})^{-2\frac{\beta(\beta-2)-1}{1+\beta}}-1}
-\frac{\zeta^{2}}{(\beta(\beta-2)-1)}(d\theta^{2}+\sin^{2}\theta d\Phi^{2}).
\end{eqnarray}
As we are interested in exploring the impact of matter terms (trace of energy-momentum tensor) involved in this theory, therefore
we will fix the value of $\beta$ and plot $\rho$ in a wider range of $\lambda$. Moreover, we choose $r_0=1$ and $\beta=0.6$ and
analyze all energy conditions. It is easy to see that these constraints are valid for $\lambda>-4\pi$ as shown in the graphs
of Figures \ref{fig7} and \ref{fig8}.

\section{Field Equations for hyperbolic wormhole space time in $f(R,T)$ Gravity}

In this section, we shall describe different cases of wormhole construction by taking hyperbolic spacetime into account.
The spacetime representing wormhole geometry in hyperbolic coordinates is defined as
\begin{equation}\label{26}
ds^2=e^{2\Phi(r)}dt^2-\frac{dr^2}{1-b(r)/r}-r^{2}(d\theta^{2}+\sinh^{2}\theta d\phi^{2}),
\end{equation}
where $\Phi$ and $b(r)$ have the usual meaning, i.e., wormhole red shift and shape functions, respectively.
From Eqs.(\ref{4}) and (\ref{26}), we obtain the following set of field equations:
\begin{eqnarray}\label{27}
\frac{b'-2}{r^2}&=&(8\pi+\lambda)\rho + \frac{\lambda}{2}(\rho-\frac{p_{r}+2p_{t}}{3}),\\\label{27*}
\frac{2\phi^{'}}{r}(1-\frac{b}{r})-\frac{b}{r^3}+\frac{2}{r^2}&=&(8\pi+\lambda)(-p_{r}) + \frac{\lambda}{2}(\rho-\frac{p_{r}+2p_{t}}{3}),\\\label{27**}
\frac{1}{2r}\bigg[\frac{1}{r}(\phi' b + b'-\frac{b}{r})+2(\phi''+(\phi')^{2})b- \phi'(2-b')\bigg]&=&(8\pi+\lambda)(-p_{t})
+ \frac{\lambda}{2}(\rho-\frac{p_{r}+2p_{t}}{3}).
\end{eqnarray}
From the above equations, the explicit form of energy density, radial and tangential pressures denoted by symbols $\rho,~ p_{r}$ and $p_{t}$,
respectively, can be obtained as follows
\begin{eqnarray}\nonumber
\rho&=&\frac{1}{12 r^{3}(4\pi+\lambda)(8\pi+ \lambda)}\bigg[(48(-2+b') \pi r + \lambda (b(2-r\phi'-2r^2(\phi^{'})^{2}+\phi^{''}))+r(b^{'}(8-r\phi^{'})+2(-10+r^2(\phi^{'})^{2}+\phi^{''}))))\bigg],\\\label{28}\\\nonumber
p_{r}&=&\frac{1}{12 r^{3}(4\pi+\lambda)(8\pi+ \lambda)}\bigg[b(48 \pi(1 + 2r \phi^{'} )
+\lambda(10+23 r \phi^{'}+2 r^2((\phi^{'})^{2}+\phi^{''})))-r(96 \pi(1+r \phi^{'})\\\label{28*}
&&+\lambda(-b^{'}(4+r\phi^{'})+2(14+12 r \phi^{'}+r^2((\phi^{'})^{2}+\phi^{''}))))\bigg],\\\nonumber
p_{t}&=&\frac{1}{12 r^{3}(4\pi+\lambda)(8\pi+ \lambda)}\bigg[-b\bigg[24\pi(-1 + r \phi^{'}+2r^{2}(\phi^{'})^{2}+\phi^{''}))+\lambda(-4+7r\phi^{'}+10r^{2}((\phi^{'})^{2}+\phi^{''})))\bigg]+\\\label{28**} &&r\bigg[-b^{'}(24\pi(1+r\phi^{'})+\lambda(2+5r\phi^{'}))+2(24r\pi(\phi^{'}+r(\phi^{'})^{2}+r\phi^{''})+
\lambda(-2+6r\phi^{'}+5r^2((\phi^{'})^{2}+\phi^{''})))\bigg]\bigg],
\end{eqnarray}
where the prime denotes the derivatives with respect to radial coordinate. It is worthy to mention here that the conservation
equation for hyperbolic spherically symmetric spacetime is same as for symmetric metric and hence given by Eq.(\ref{8}). It is
easy to check that we have system of three equations with five unknown functions. To develop solutions representing hyperbolic
wormholes, we have to fix two of these unknowns. In this work, we will solve these equations by considering the following three
possibilities:
\begin{itemize}
\item EoS for radial or tangential pressure;
\item Condition for isotropic pressure, i.e., $p_r=p_t$;
\item Some known and interesting choices for $\phi(r)$ and $b(r)$.
\end{itemize}
In the following subsections, we shall discuss these cases separately and present the graphical analysis of obtained solutions.

\subsection{Hyperbolic Wormhole with Equation of State in $f(R,T)$ Gravity}

\begin{figure}
\centering
\includegraphics[width=0.46\textwidth]{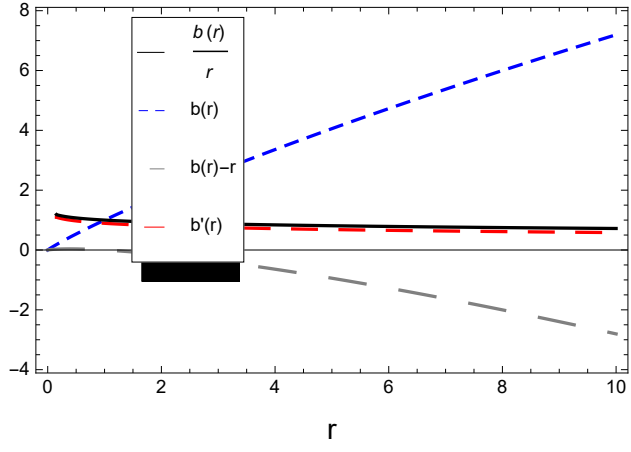}
\caption{\scriptsize{Evolution of shape function given by Eq.(ref{31}) to check its consistency with the axioms.
Here, shape function (blue dashes), $b'(r)$ (red dashes), $b(r)-r$ (gray dashes) and $\frac{b(r)}{r}$ (black), where the
chosen parameters are $\omega=-16, ~\lambda=-15$ and $r_0=1$.}}
\label{fig9}
\end{figure}

In this subsection, we shall try to explore the existence of pseudospherical wormhole solutions filled by the matter
with anisotropic pressure by taking the following EoS for the radial pressure into account:
\begin{eqnarray}\label{29}
p_r=\omega\rho.
\end{eqnarray}
From Eq.(\ref{28}) and the above EoS parameter, it can be written as
\begin{eqnarray}\nonumber
&&\frac{1}{r (4\pi + \lambda)(8\pi+ \lambda)}\bigg[-b (48 \pi (1 + 2 r \phi^{'})+\lambda(-r \phi^{'}(-23 + \omega)-2(-5 + \omega)
+2r^2((\phi^{'})^2 + \phi^{''})(1 + \omega)))\\\nonumber
&&+r (48 \pi(2 + 2 r\phi^{'}+(-2 + b^{'})\omega)+\lambda(-b^{'}(4-8\omega+r \phi^{'}(1+\omega))
+2 (14 + 12 r \phi^{'}-10 \omega+ r^2 (\phi^{'})^2 + \phi^{''})(1 + \omega)))))\bigg]=0.
\end{eqnarray}
For zero tidal force wormhole, one needs to impose the condition $\phi(r)=0$ and consequently, we obtain
\begin{eqnarray}\label{31}
b(r)= 2r + A (2r (\lambda - 12 \pi \omega - 2 \lambda \omega))^{\frac{24 \pi - \lambda (-5 + \omega)}{24 \pi \omega + \lambda (-2 + 4 \omega)}},
\end{eqnarray}
where $A$ is an integration constant. For this shape function, the graphical illustration of different axioms which are necessary for
a viable wormhole geometry is provided in Figure \ref{fig9}.
\begin{figure}
\centering
\includegraphics[width=0.46\textwidth]{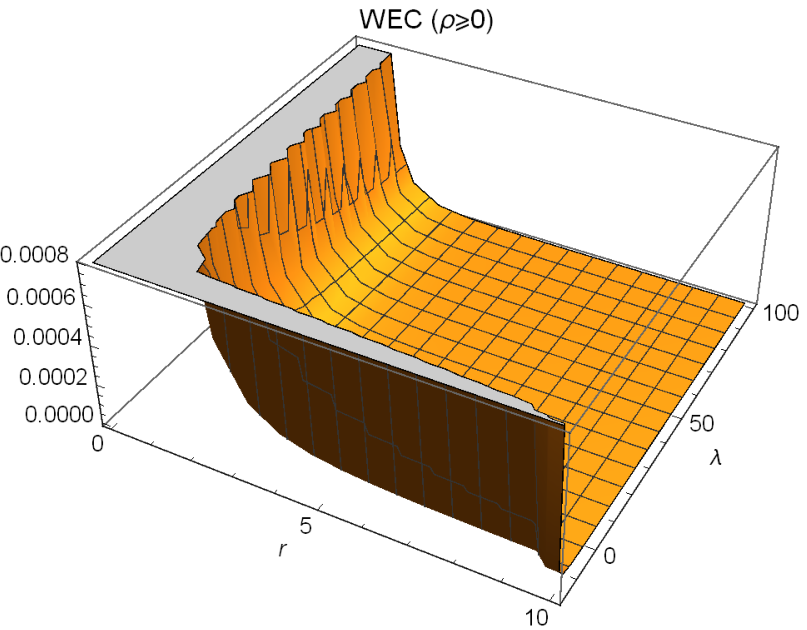}
\includegraphics[width=0.46\textwidth]{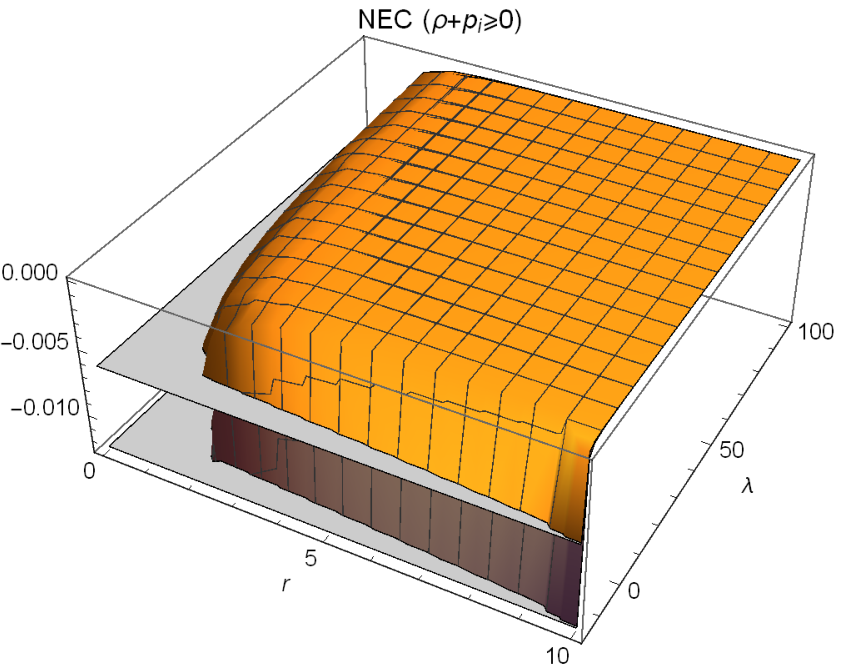}
\caption{\scriptsize{Evolution of energy density and NEC ($\rho+p_i$) versus $r$. In the left panel, the plot indicates the behavior
of energy density. In the right panel, the gray plot shows the behavior of NEC ($\rho+p_r$) and orange plot shows the behavior of
NEC ($\rho+p_t$), where $\omega=-16$ and $r_0=1$.}}
\label{fig10}
\end{figure}
\begin{figure}
\centering
\includegraphics[width=0.46\textwidth]{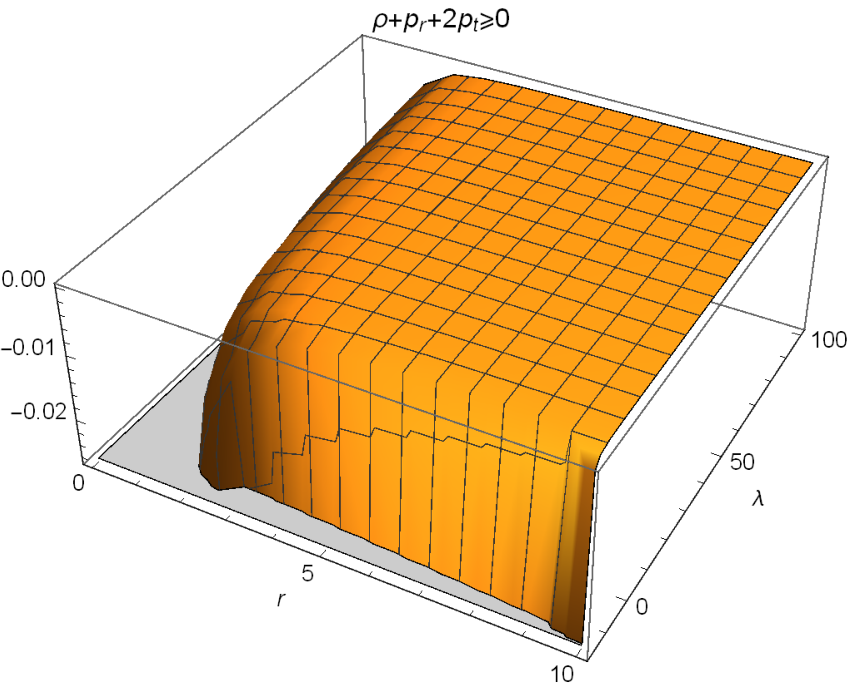}
\includegraphics[width=0.46\textwidth]{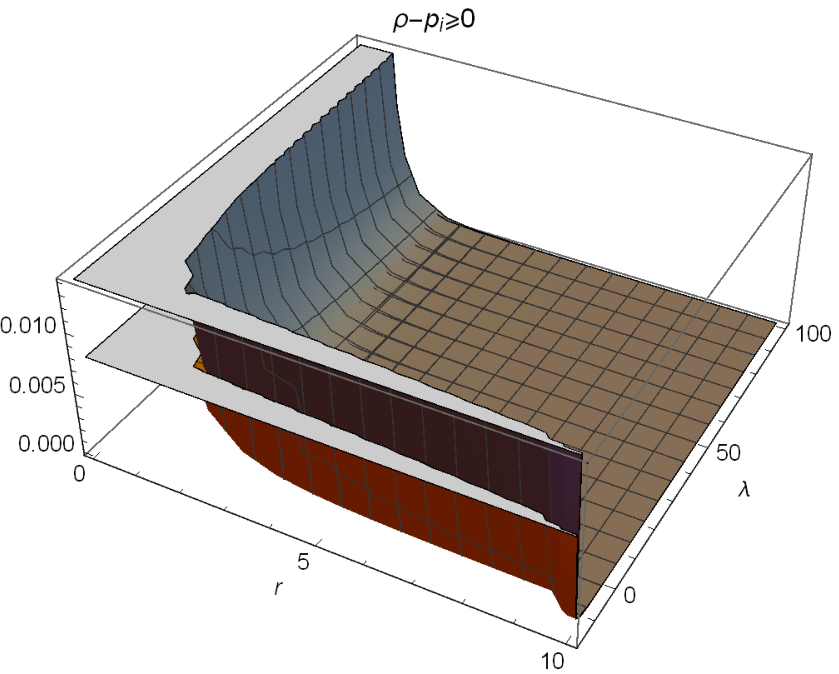}
\caption{\scriptsize{Evolution of SEC ($\rho+p_r+2p_t$) and DEC ($\rho+p_i$) versus $r$. In the left panel, the plot indicates
the behavior of SEC. In the right panel, the gray plot shows the behavior of DEC ($\rho-p_r$) and orange plot shows the behavior
of DEC ($\rho-p_t$) for $\omega=-16$ and $r_0=1$.}}\label{fig11}
\end{figure}
Consequently, one may write wormhole metric, energy density and tangential pressure as follows
\begin{eqnarray}\label{33}
ds^2&=&dt^2 - \frac{dr^2}{(\frac{r}{r_o})^{-\frac{24\pi(\omega-1)+\lambda(7-5\omega)}{24\pi\omega-2\lambda+4\lambda\omega}}-1}
-r^2(d\theta^2+\sinh^2\theta d\phi^2),\\\label{33*}
\rho&=&-\frac{3r_0(\frac{r}{r_o})^{\frac{24\pi-\lambda(-5+\omega)}{24\pi\omega-2\lambda+4\lambda\omega}}}{r^3(24\pi\omega-2\lambda+4\lambda\omega)},\\\label{33**}
p_t&=&-\frac{3r_0(-1+\omega)(\frac{r}{r_o})^{\frac{24\pi-\lambda(-5+\omega)}{24\pi\omega-2\lambda+4\lambda\omega}}}{ 2r^3 (24\pi\omega-2\lambda+4\lambda\omega)}.
\end{eqnarray}
Next we shall describe the energy conditions of wormhole matter content graphically with respect to radial coordinate.
These graphs are provided in Figures \ref{fig10} and \ref{fig11}. The left panel of Figure \ref{fig10} provides the validity of WEC,
i.e., $\rho\geq0$ which indicates that energy density remains positive for $\lambda<0$. From the right panel of same graph, it
is easy to observe that NEC for both radial and tangential pressures remain invalid. Likewise, the left and right graphs of Figure
\ref{fig11} shows that SEC violates while DEC remains valid for both radial and tangential pressures. Thus, it can be concluded that
the WEC and DEC are valid in the present setup whereas the SEC and NEC are incompatible for the choice $\lambda<0$.
\begin{figure}
\centering
\includegraphics[width=0.46\textwidth]{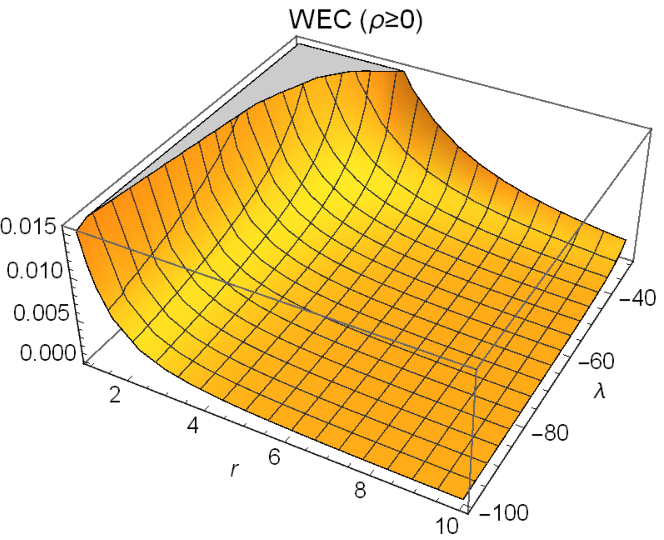}
\includegraphics[width=0.46\textwidth]{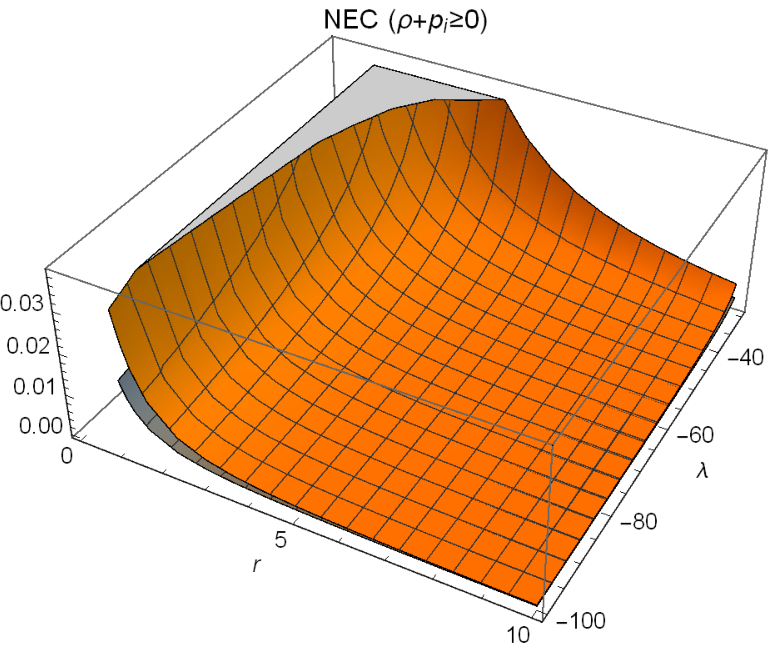}
\caption{\scriptsize{Evolution of energy density ($\rho$) and NEC ($\rho+p_i$) versus $r$. In the left panel, the right plot
represents the behavior of energy density ($\rho$). In the right panel, the orange plot shows the behavior of $\rho+p_r$ and
gray plot represents the behavior of $\rho+p_t$ for $\alpha=0.5$ and $r_0=1$.}}
\label{fig12}
\end{figure}

\subsection{Zero tidal force Hyperbolic Wormholes along with Isotropic fluid or a Power Law Shape Function in $f(R,T)$ Gravity}

\begin{figure}
\centering
\includegraphics[width=0.46\textwidth]{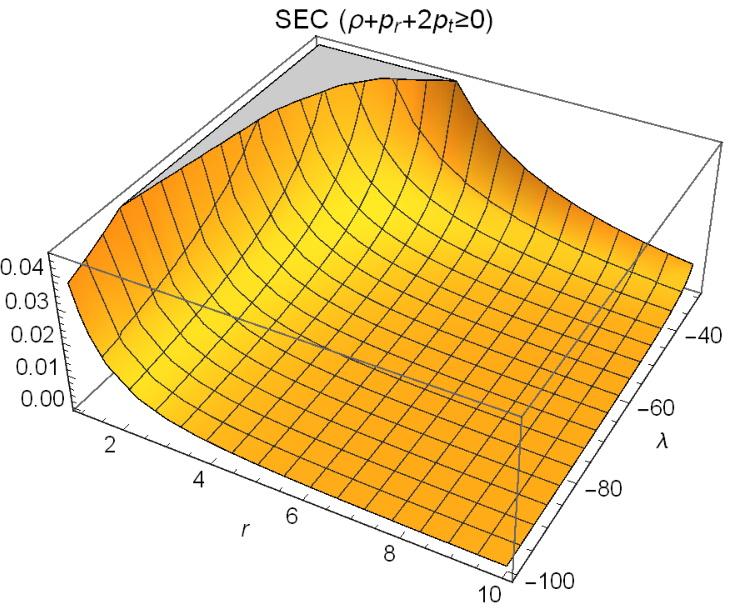}
\includegraphics[width=0.46\textwidth]{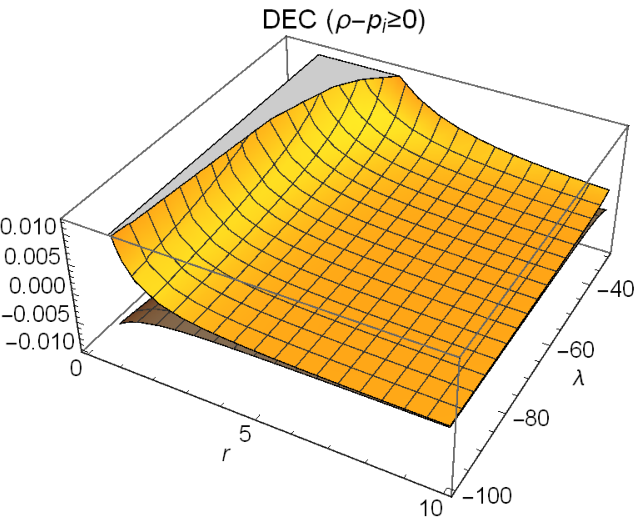}
\caption{\scriptsize{Evolution of SEC ($\rho+p_r+2p_t$) and DEC ($\rho-p_i$) versus $r$. In the left panel,
the plot shows the behavior of SEC. In the right panel, gray plot represents the behavior of DEC ($\rho-p_r$)
and yellow plot represents the behavior of DEC ($\rho-p_t$) with $\alpha=0.5$ and $r_0=1$.}}
\label{fig13}
\end{figure}

In this section, we shall construct wormhole geometry by considering two cases. In the first case, we shall construct M-T hyperbolic
wormhole by taking an interesting choice of shape function given by $b(r)=r_0(\frac{r}{r_0})^{\alpha}$ along with zero red shift function,
i.e., $\phi(r)= 0$. The graphical illustration of this shape function has already been discussed in the previous section which
indicated that it is viable choice for $\alpha<0$. It is worthy to mention here that in case of hyperbolic geometry, the behavior
of this shape function is accurate and consistent with all axioms for $\alpha>0$. Inserting the value of red shift as well as
shape function in Eq.(\ref{29}), we obtain the metric, $\rho$ and $p$ in the following form:
\begin{eqnarray}\label{34}
ds^2&=&dt^2-\frac{dr^2}{1-(\frac{r}{r_o})^{\alpha -1}}-r^2(d\theta^2+\sinh^2\theta d\phi^2),\\\label{34*}
\rho&=&\frac{\pi(-48r+24\alpha r_0(\frac{r}{r_o})^{\alpha})+(-10r+(1+4\alpha)r_0(\frac{r}{r_0})^{\alpha})\lambda}{6r^3(4\pi+\lambda)(8\pi+\lambda)},\\\label{34**}
p_t&=&\frac{\pi(-24\pi-4\lambda) r_0((\frac{r}{r_o})^{\alpha})+r(-4\lambda+(24\pi+2\lambda)\alpha(\frac{r}{r_0})^{\alpha})}{12r^3(4\pi+\lambda)(8\pi+\lambda)}.
\end{eqnarray}
For the obtained solutions, we plot the models of all energy conditions versus radial coordinates as shown in Figures \ref{fig12} and
\ref{fig13}. From these graphs, it can be seen that all energy conditions are valid for $\lambda<-30$ except the DEC ($\rho-p_r$) for
radial pressure.
In the second case, we will construct new pseudospherical wormholes with isotropic pressure by imposing the condition given by
$p_r=p_t$. For hyperbolic spacetime, using Eq.(\ref{28}), we obtain the following differential equation:
\begin{eqnarray}\label{38}
\phi^{''}+\phi^{'^{2}}+\frac{(b^{'}r-3r+5b)\phi^{'}}{2r(b-r)}=\frac{-b^{'}r-b+4r}{2r^{2}(b-r)}
\end{eqnarray}
which must be satisfied by the shape and red shift functions. This differential equation can be re-arranged and the shape function
can be determined as
\begin{eqnarray}\label{39}
b(r)=\bigg(\int\frac{2(2+3r\phi^{'}+r^2(\phi^{'2}+\phi^{''}))e^{-\int\frac{1+2r^{2}(\phi^{'2}+\phi^{''})+5r\phi^{'}}{r(1+r\phi^{'})}dr}}{1+r\phi^{'}}dr+B\bigg)\times
e^{-\int\frac{1+2r^{2}(\phi^{'2}+\phi^{''})+5r\phi^{'}}{r(1+r\phi^{'})}dr},
\end{eqnarray}
where $B$ is an integration constant. Now we have two set of differential equations with four unknowns namely
shape function, red shift function, density and pressure. For solving these equations, one should use some valid
assumptions. In the present work, we pick an interesting choice of red shift function given by $\phi(r)=\phi_0$=constant (zero tidal force).
By inserting this condition in Eq.(\ref{39}), we obtain a spacetime with constant curvature defined by $b(r)=2r+\frac{B}{r}$.
In this case, the wormhole spacetime, energy density and isotropic pressure can be written as
\begin{eqnarray}\nonumber
ds^2&=&dt^2-\frac{dr^2}{(\frac{r}{r_0})^{-2}-1}-r^2(d\theta^2+\sinh^2\theta d\phi^2),\\\nonumber
\rho&=&\frac{(r_0)^2}{2r^4(4\pi+\lambda)},\\\nonumber
p_r&=&p_t=-\frac{(r_0)^2}{2r^4(4\pi+\lambda)}.
\end{eqnarray}
Also, by evaluating the flaring out condition at throat, we obtain $b'(r_0)=2\nless 1$ which shows that no zero tidal force
hyperbolic wormhole exists in $f(R,T)$ gravity.

\subsection{Hyperbolic Wormhole with Non-vanishing RedShift Function and Isotropic Pressure in $f(R,T)$}

\begin{figure}
\centering
\includegraphics[width=0.46\textwidth]{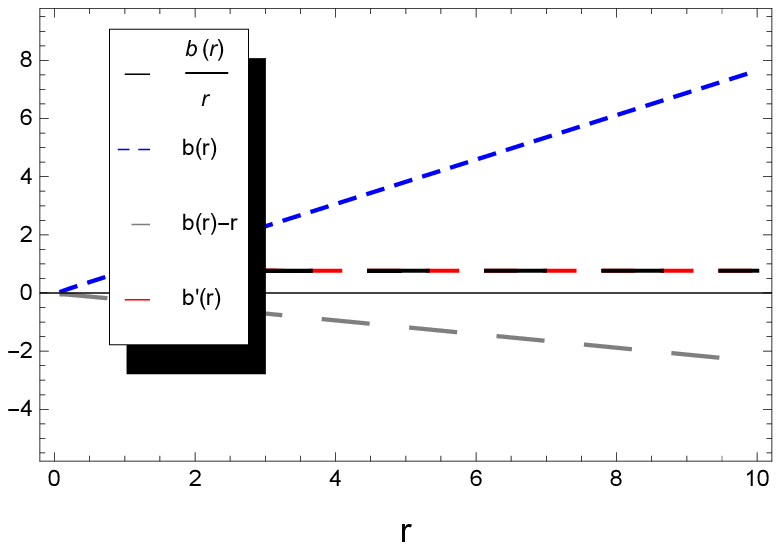}
\includegraphics[width=0.46\textwidth]{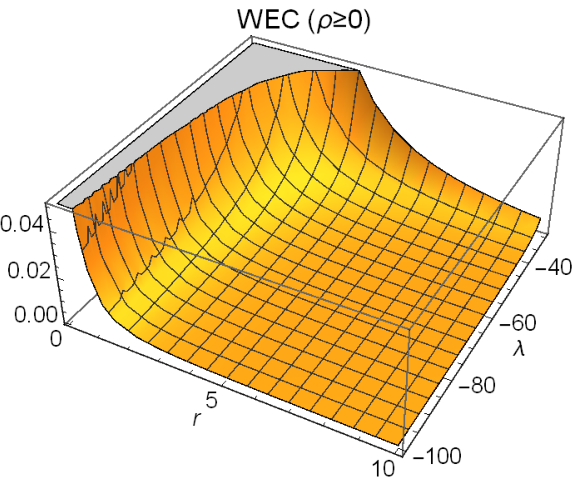}
\caption{\scriptsize{Evolution of shape function satisfying all axioms and energy density versus $r$. In the left panel,
shape function (blue plot), $b'(r)$ (red plot), $b(r)-r$ (gray plot) and $\frac{b(r)}{r}$ (black plot). In the right panel,
the plot represents the behavior of energy density where the chosen free parameters are $r_0=.1$ and $\beta=3.5$.}}\label{fig16}
\end{figure}

\begin{figure}
\centering
\includegraphics[width=0.46\textwidth]{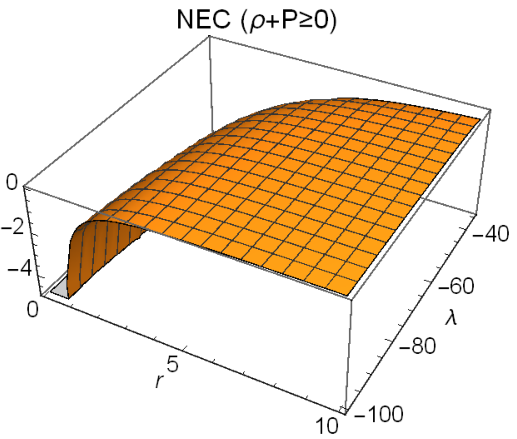}
\includegraphics[width=0.46\textwidth]{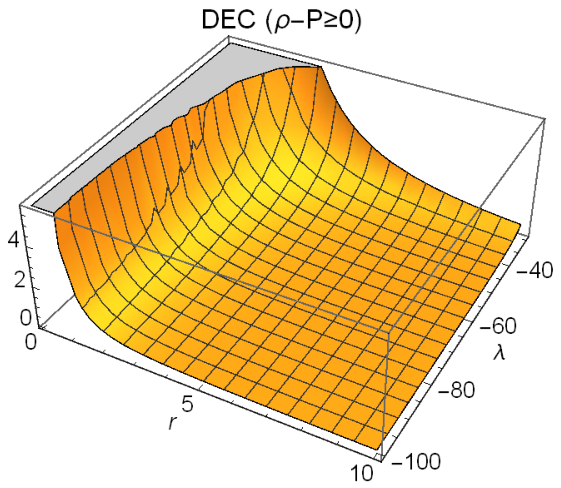}
\caption{\scriptsize{Evolution of NEC ($\rho+p_i$) and DEC ($\rho-p_i$) versus $r$. In the left panel,
$\rho+p_r$ (yellow plot) and $\rho+p_t$ (gray plot) while, in the right panel, $\rho-p_r$ (yellow plot) and
$\rho-p_t$ (gray plot) for $r_0=.1$ and $\beta=3.5$.}}
\label{fig17}
\end{figure}

\begin{figure}
\centering
\includegraphics[width=0.46\textwidth]{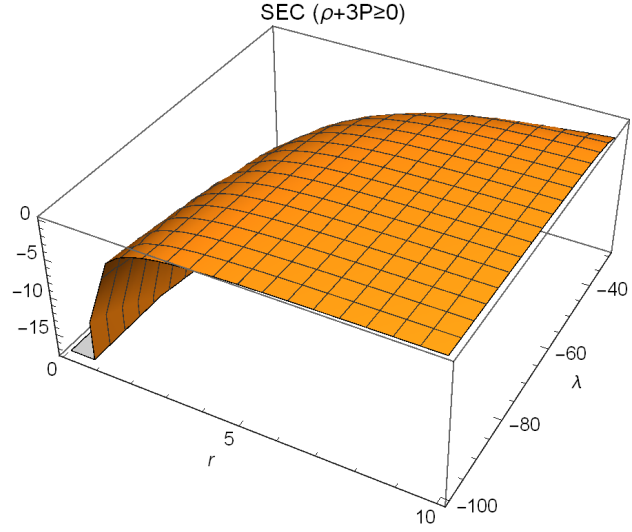}
\caption{\scriptsize{Evolution of SEC ($\rho+p_r+2p_t$) versus radial coordinate.}}
\label{fig18}
\end{figure}

Here we shall consider an interesting power law form of red shift function given by $e^{\phi(r)}=(\frac{r}{r_0})^{\beta}$
with $\beta$ as some arbitrary constant. By inserting the value of $e^{\phi(r)}$ in Eq.(\ref{38}), we obtain the
following form of shape function given by
\begin{eqnarray}\label{35}
b(r)=\frac{\beta^{2}-2\beta-2}{\beta(\beta-2)-1}r - cr^{-\frac{2\beta^{2}-5\beta-3}{1+\beta}},
\end{eqnarray}
where $c$ is a constant of integration. The graphical behavior of this shape function is shown in Figure \ref{fig16} which
indicates that all axioms, essential for a valid shape function, are satisfied.
By imposing, the wormhole condition $b(r_{0})=r_{0}$, Eq.(\ref{35}) leads to the following expression for wormhole metric, $\rho$ and $p$:
\begin{eqnarray}\nonumber
ds^{2}&=&(\frac{r}{r_{0}})^{2\beta}dt^{2}-\frac{(\beta(\beta-2)-1)}{(\frac{r}{r_{0}})^{-2\frac{\beta(\beta-2)-1}{1+\beta}}-1}dr^{2}
-r^{2}(d\theta^{2}+\sinh^{2}\theta d\phi^{2}),\\\nonumber
\rho&=&\frac{1}{6 r^{3}(1+\beta)(-1-2\beta+\beta^2)(8\pi+\lambda)(4\pi+\lambda)}\bigg[(24\pi(r\beta(2+\beta+\beta^2)+(\frac{r}{ro})^{(3+5\beta-2 \beta^2)/(1+\beta)}r_0(-3-5\beta+2 \beta^2))\\\nonumber
&+&(r\beta(9+5\beta-4\beta^2)+(\frac{r}{ro})^{(3+ 5\beta-2\beta^2)/(1 + \beta)}r_0(-13-21\beta+10\beta^2)\lambda)\bigg],\\\nonumber
p_r&=&\frac{1}{6 r^{3}(1+\beta)(-1 - 2 \beta + \beta^2)(8\pi+ \lambda)(4\pi+\lambda)}\bigg[(24\pi(1+\beta)(r\beta^2+(\frac{r}{ro})^{(3+5\beta-2\beta^2)/(1+\beta)}r_0(1+2 \beta))\\\nonumber
&+&(r\beta(3+7\beta+4\beta^2)+(\frac{r}{ro})^{(3+5\beta-2\beta^2)/(1+\beta)}r_0(1+9\beta+14\beta^2)\lambda)\bigg].
\end{eqnarray}
By the re-scaling $\zeta^{2}=(\beta(\beta-2)-1)r^{2}$ of the above wormhole metric, it can be written as
\begin{eqnarray}\label{25}
ds^{2}=(\frac{\zeta}{\zeta_{0}})^{2\beta}dt^{2}-\frac{d\zeta^2}{(\frac{\zeta}{\zeta_{0}})^{-2
\frac{\beta(\beta-2)-1}{1+\beta}}-1}
-\frac{\zeta^{2}}{(\beta(\beta-2)-1)}(d\theta^{2}+\sinh^{2}\theta d\phi^{2}).
\end{eqnarray}
Since our main objective is to explore the impact of matter terms in the considered gravitational theory, therefore
we fix the value of $\beta$ and plot energy density by varying the value of $\lambda$. Moreover, we choose $r_0=1$ and $\beta=3.5$.
We analyze the energy conditions graphically for the obtained model. From the right graph of Figure \ref{fig16}, it is seen that
the energy density remain positive versus radial coordinate. From the graphs of Figure \ref{fig17}, it can be observed that the
NEC ($\rho +p_r\geq0$) is valid for $\lambda<-35$ while for tangential pressure, it is invalid. From the same Figure, it is seen that
the validity of DEC ($\rho-p_i\geq0$) is guaranteed for tangential pressure, whereas SEC violates for $\lambda <-35$ as shown in Figure
\ref{fig18}).

\section{Conclusion and Final Remarks}

The construction of viable wormhole geometries in modified gravity theories has always been a fascinating topic for researchers.
In the present work, we propose interesting exact solutions for both classes of static spherically symmetric and pseudospherically
symmetric spacetimes representing wormholes in the well-famed $f(R,T)$ gravitational framework. The considered gravitational theory is an
interesting choice in this respect as it allows the interaction of curvature and matter. For the sake of simplicity, we have
considered linear model of $f(R,T)$ function representing minimal interaction of matter and curvature. Our primary objective is to check
what ranges of coupling parameter $\lambda$, present in $f(R,T)$ function, can allow the existence of wormholes as well as the validity
of energy condition bounds. In this work, we have obtained some specific solutions either by considering EoS or
by assuming some interesting choices of red shift or shape functions. The obtained solutions are then examined graphically and
the validity of energy condition bounds have been explored in each case. It is seen that for all obtained spherically symmetric wormholes,
we have positive energy density at the throat and radial pressure is negative for $\lambda>-4\pi$ and $\lambda<-8\pi$ satisfying
all axioms for shape function. For pseudospherically symmetric wormhole, energy density is positive for all conditions and radial pressure
may be positive or negative for different ranges of $\lambda$.

For zero tidal force wormhole one needs to assume $\phi(r)=$constant, and we have discussed two possibilities. Firstly, we obtain shape
function using the main condition for perfect fluid, i.e., $p_r=p_t$. Secondly, we have imposed a restricted choice of shape function
for both spherically and hyperbolic symmetric metrics. In both cases, energy density remains positive and radial pressure is negative
for $\lambda>-4\pi$, but for pseudospherically symmetric wormhole, the energy density is also positive and radial pressure is negative
for $\lambda>-4\pi$ in first possibility. But for the restricted choice of shape function, the energy density and pressure are positive
for $\lambda\leq-10\pi$.

We have also discussed a possibility with isotropic fluid using the linear EoS. We have obtained the shape function by imposing zero-tidal force
condition for red shift function in both symmetric and pseudospherically symmetric wormhole cases by taking linear EoS into account. We have
observed that energy density is positive and radial pressure is negative for $\lambda>-4\pi$ for spherically symmetric wormhole. Whereas,
for hyperbolic wormhole energy density remains positive and pressure turns out as negative for $\lambda<0$. Lastly, we have imposed a
restricted choice of red shift function to evaluate the most general form of $b(r)$ using isotropic pressure condition. For this condition,
in case of symmetric wormhole, same results have been obtained for energy density and radial pressure for $\lambda>-4\pi$. In case of
hyperbolic wormhole, both energy density and radial pressure are positive for $\lambda\leq -12\pi$.

At the end, we have applied the energy conditions to material content of symmetric and pseudospherically symmetric wormhole solutions for all
restricted choices of red shift and shape functions as well as linear equation of state with isotropic pressure. One can easily observe from all
figures that either NEC, DEC and SEC are violated or valid for different ranges of $\lambda$ while WEC is always valid for both symmetric and
pseudospherically symmetric wormholes. A brief comparison of both classes of wormholes is listed in Table \ref{Table1}.
In this table, we indicate that which energy condition is satisfied or remained invalid corresponding to specified wormhole geometry, which can be explained as follows:
\begin{table}
\centering
\begin{tabular}{|c|c|c|c|c|c|}
  \hline
  Metric & Constraints& WEC ($\rho\geq0$) & NEC ($\rho+p_r\geq0$) & DEC ($\rho-p_r\geq0$)& SEC ($\rho+ p_r$ \\
         &             &                  &      ($\rho+p_t\geq0$) &     ($\rho-p_t\geq0$)&  $+2p_t\geq0$)  \\\hline
% $ds^2=dt^2-\frac{dr^2}{1-(\frac{r}{r_0})^2}-r^2d\Omega^2$ &$\phi(r)$=constant, & Yes & Yes & Yes& Yes\\
%                                                           & $b(r)=\frac{r^3}{r_0^2}$, $\lambda>-4\pi$ & & & &\\
%                                                           & $p_r=p_t$ & & & & \\\hline
$ds^2=dt^2-\frac{dr^2}{1-(\frac{r}{r_0})^{\alpha-1}}-r^2d\Omega^2$ &$\phi(r)$=constant, & Yes&No &Yes & Yes\\
                                                                   &$b(r)=r_0(\frac{r}{r_0})^{\alpha}$, & &Yes & No&\\
                                                                   &$\lambda<-8\pi$& & & & \\\hline
 $ds^2=dt^2-\frac{dr^2}{(\frac{r}{r_0})^{\frac{-(1+\omega)(\lambda_1)}{\lambda_2}}-1}$ & $\phi(r)=0$, &Yes & No & Yes & No \\
$-r^2d\Omega^2$, $\lambda_1=12\pi+2\lambda$ &$b(r)=r_0(\frac{r}{r_0})^{\frac{-3\lambda_3}{\lambda_2}}$, & & No& Yes & \\
$\lambda_2=12\pi\omega-\lambda+2\omega\lambda$& $\lambda_3=4\pi+\lambda$, & & & & \\
                                                & $p_r=\omega\rho$, & & & & \\
                                                & $\lambda>-4\pi$ & & & & \\\hline
$ds^2=(\frac{r}{r_0})^{2\beta}dt^2-\frac{\beta_1dr^2}{(\frac{r}{r_0})^{\frac{-2\beta_1}{1+\beta}}-1}$ &$\phi(r)=\ln(\frac{r}{r_0})^{2\beta}$,   & Yes  & Yes & Yes & No        \\
$-r^2d\Omega^2$, $\beta_1=\beta^2-2\beta-1$ & $b(r)=\frac{(\beta^2-2\beta)r}{\beta_1}$  &  & Yes & Yes & \\
                                             & $-\frac{r_0}{\beta_1}(\frac{r}{r_0})^{\frac{-(2\beta + 1)*(\beta - 3)}{1+\beta}}$, & & & & \\
                                             & $\lambda>-4\pi$ & & & & \\\hline
$ds^2=dt^2-\frac{dr^2}{(\frac{r}{r_0})^{-\frac{\varsigma_2}{\varsigma_1}}-1}-$ &$\phi(r)=0$, &Yes &No &Yes &No \\
$r^2d\Omega_{ps}^2$, $\varsigma_1=24\pi\omega-2\lambda+4\lambda\omega$ &$b(r)=2r-r_0(\frac{r}{r_0})^{\frac{\varsigma_3}{\varsigma_1}}$, & &No &Yes & \\
$\varsigma_2=-24\pi(\omega-1)+\lambda(7-5\omega)$ & $\varsigma_3=24 \pi + \lambda (-5 + \omega)$, & & & & \\
                                                  & $p_r=\omega\rho$,  & & & & \\
                                                  & $\lambda<0$ & & & & \\\hline

 $ds^2=dt^2-\frac{dr^2}{1-(\frac{r}{r_0})^{\alpha-1}}-$ &$\phi(r)=0$, &Yes &Yes &No & Yes\\
 $r^2d\Omega_{ps}^2$ &$b(r)=r_0(\frac{r}{r_0})^{\alpha}$ & & & & \\
                        & $\lambda\leq-10\pi$ & & & & \\\hline
 %$ds^2=dt^2-\frac{dr^2}{(\frac{r}{r_0})^{-2}-1}-r^2d\Omega_{ps}^2$ &$\phi(r)$=constant, & Yes & No & Yes& No\\
%                                                           & $b(r)=2r-\frac{r_0^2}{r}$, & & & &\\
%                                                           & $p_r=p_t$, $\lambda>-4\pi$ & & & & \\\hline
 $ds^2=(\frac{r}{r_0})^{2\beta}dt^2-\frac{\beta_1dr^2}{(\frac{r}{r_0})^{\frac{-2\beta_1}{1+\beta}}-1}$ &$\phi(r)=\ln(\frac{r}{r_0})^{2\beta}$,   & Yes  & Yes & No & No        \\
$-r^2d\Omega_{ps}^2$, $\beta_1=\beta^2-2\beta-1$ & $b(r)=\frac{(\beta^2-2\beta-2)r}{\beta_1}$  &  & No & Yes & \\
                                             & $-\frac{r_0}{\beta_1}(\frac{r}{r_0})^{\frac{-(2\beta + 1)*(\beta - 3)}{1+\beta}}$, & & & & \\
                                             & $\lambda\leq-12\pi$ & & & & \\\hline
\end{tabular}
\caption{In this table, we have listed all obtained wormhole solutions. It indicates that which of the energy conditions, i.e.,
WEC, NEC, DEC and SEC are valid. It is worthy to mention here that for wormhole with isotropic pressure, the SEC becomes $\rho+3p$.
Also, here $d\Omega^2=d\theta^2+r^2\sin^2\theta d\Phi^2$ and $d\Omega_{ps}^2=d\theta^2+r^2\sinh^2\theta d\Phi^2$.} \label{Table1}
\end{table}

\begin{enumerate}
\item In case of spherical wormhole,
\begin{itemize}
\item when $\phi(r)$ is constant and $b(r)=r_0(\frac{r}{r_0})^{\alpha}$, the WEC, SEC, NEC with transverse pressure
and DEC with radial pressure are valid for $\lambda<-8\pi$, while the NEC with radial pressure and DEC with transverse pressure
are not satisfied for $\lambda<-8\pi$.
\item For $\phi(r)=0$ and $b(r)=r_0(\frac{r}{r_0})^{\frac{-3(4\pi+\lambda)}{12\pi\omega\lambda+2\omega\lambda}}$, the WEC and
DEC are satisfied for $\lambda>-4\pi$, while the NEC and SEC are not satisfied for $\lambda>-4\pi$.
\item For $\phi(r)=\ln(\frac{r}{r_0})^{2\beta}$ and $b(r)=\frac{\beta^2-2\beta}{\beta^2-2\beta-1}r-\frac{r_0}{\beta_1}(\frac{r}{r_0})^{\frac{-(2\beta+1)(\beta-3)}{1+\beta}}$,
all energy constraints are valid except SEC for $\lambda>-4\pi$.
\end{itemize}
\item In case of hyperbolic wormhole,
\begin{itemize}
\item when $\phi(r)=0$ and $b(r)=2r-r_0(\frac{r}{r_0})^{\frac{24\pi-5\lambda+\omega\lambda}{24\pi\omega-2\lambda+4\lambda\omega}}$,
the WEC and DEC are valid for $\lambda<0$, while the NEC and SEC are not valid for $\lambda<0$.
\item For $\phi(r)=0$ and $b(r)=r_0(\frac{r}{r_0})^{\alpha}{12\pi\omega\lambda+2\omega\lambda}$, the WEC, NEC and SEC are valid,
while the DEC is violated for $\lambda\leq-10\pi$.
\item For $\phi(r)=\ln(\frac{r}{r_0})^{2\beta}$ and $b(r)=\frac{\beta^2-2\beta-2}{\beta^2-2\beta-1}r-\frac{r_0}{\beta_1}(\frac{r}{r_0})^{\frac{-(2\beta+1)
    (\beta-3)}{1+\beta}}$,
the WEC, the NEC with radial pressure and the DEC with transverse pressure are satisfied for $\lambda\leq-12\pi$, while the SEC and NEC, and
DEC with other components of pressure are not satisfied for $\lambda\leq-12\pi$.
\end{itemize}
\end{enumerate}
Many authors have studied the topic of wormholes existence in the framework of
$f(R,T)$ theory. In this context, Sahoo and his collaborators \cite{8} studied the spherically symmetric wormhole solution for
phantom case, i.e., $\omega<-1$ in $f(R,T)$ modified gravity. They adopted the EoS given by $p_r=\omega\rho$ to obtain the form of shape function
which can satisfy all the necessary axioms for the existence of wormhole geometry and also examined corresponding energy condition bounds.
Similarly, Moraes and Sahoo \cite{9} proposed the wormhole solution by applying the power law form of shape function and analyzed
all energy constraints in exponential $f(R,T)$ gravity. In another study, Zubair et al. \cite{21} proposed wormhole
geometries by adopting simple linear as well as cubic forms of $f(R,T)$ function and further they considered the non-commutative geometrical
aspects of string theory in the context of $f(R,T)$ gravity. In a paper \cite{z32}, the researchers have explored the idea of viable charged
wormhole solutions in $f(R,T)$ gravity. They have assumed simple linear generic model given by $f(R,T)=R+2T$ along with the
ordinary matter as the total pressure of anisotropic fluid. Further, the existence of static wormhole model have been explored by
by some researchers where they utilized different kind of shape functions \cite{z33}. In another paper \cite{z34}, authors have investigated
the wormhole modeling by considering a specific general shape function in the quadratic $f(R,T)$ gravity. In a recent papers \cite{z35},
Banerjee and his collaborator adopted different strategies to construct wormhole geometry using isotropic pressure in $f(R,T)$ theory and
concluded that all energy constraints are valid for the proposed models, while Bhar and his collaborators \cite{Bhar}, proposed the M-T
wormhole solution which admits conformal motion in $f(R,T)$ gravity. They employed the phantom energy EoS, i.e., $p_r=\omega \rho$,~~$\omega<-1$
to constraint their model and showed that wormhole solution exist for both positive and negative values of coupling constant $\lambda$.

In majority of the previously listed works, authors have considered a specific form of shape function
and analyze the possibility of static and spherically symmetric wormhole existence satisfying energy condition bounds. In the present work,
we have explored both the spherically symmetric and pseudospherically symmetric wormhole solutions for M-T spacetime in $f(R,T)$ theory.
The present work can be considered as an extension of the papers \cite{3} and \cite{4}, where authors have discussed similar cases of
wormhole existence by using spherically and pseudospherically symmetric spacetimes in the framework of GR. In this work, we have not only
checked the validity of energy bounds by taking some type of shape functions into account but also explored the possible forms of $\phi(r)$
and $b(r)$ functions by taking the matter with EoS parameter for $p_r$ or $p_t$ through $f(R,T)$ field equations. Further, we have presented
the validity of energy conditions bounds for both spherical and hyperbolic wormholes in the presence of single perfect fluid, i.e., a source of matter
with isotropic pressure with a wider range of parameter $\lambda$. Up to the best of our knowledge, the only wormhole solution
discussed in literature are non-asymptotically flat wormhole with isotropic pressure as pointed out in the References \cite{28}-\cite{30}.
In the present work, we have discussed the asymptotically flat wormholes which may or may not satisfy all energy constraints
with a wider range of $\lambda$ for both types of wormhole spacetimes.

\vspace{.5cm}

\section*{Acknowledgments}

M. Zubair thank the Higher Education Commission, Islamabad, Pakistan for its
financial support under the NRPU project with grant number
$\text{5329/Federal/NRPU/R\&D/HEC/2016}$.

\end{document}